\documentclass[aps,prd,twocolumn,floatfix,superscriptaddress]{revtex4-1}
\usepackage{tabularx}
\usepackage{graphicx}
\usepackage{amsmath}
\usepackage{amssymb}
\usepackage{comment}
\usepackage[dvipsnames]{xcolor}
\usepackage{xspace}
\usepackage{multirow}
\usepackage{natbib}
\usepackage{hyperref}
\usepackage{url}
\newcommand{\nruns}{N_{\rm runs}}
\newcommand{\niter}{N_{\rm iter}}

\DeclareMathOperator*{\argmax}{argmax}

\newcommand{\algoname}{SAPTARISHI\xspace}
\newcommand{\fstat}{$\mathcal{F}$-statistic\xspace}
\newcommand{\isis}{ISE\xspace}
\newcommand{\visis}{vanilla-\isis\xspace}
\newcommand{\xMBLT}{xBSE\xspace}
\newcommand{\iMBLT}{iBSE\xspace}

\newcommand{\Np}{N_{\rm p}}
\newcommand{\marg}{\rm marg}

\begin{document}

\title{Iterative time-domain method for resolving multiple gravitational wave sources in Pulsar Timing Array data}

\author{Yi-Qian Qian}
\email{yiqian@hust.edu.cn}
\affiliation{MOE Key Laboratory of Fundamental Physical Quantities Measurements, 
Hubei Key Laboratory of Gravitation and Quantum Physics, PGMF, 
Department of Astronomy and School of Physics, Huazhong University of Science and Technology, 
Wuhan 430074, China}

\author{Soumya D.~Mohanty}
\email{soumya.mohanty@utrgv.edu}
\affiliation{Department of Physics and Astronomy, University of Texas Rio Grande Valley, One West University Blvd., Brownsville, Texas 78520, USA}

\author{Yan Wang}
\email{ywang12@hust.edu.cn}
\affiliation{MOE Key Laboratory of Fundamental Physical Quantities Measurements, 
Hubei Key Laboratory of Gravitation and Quantum Physics, PGMF, 
Department of Astronomy and School of Physics, Huazhong University of Science and Technology, 
Wuhan 430074, China}

\begin{abstract}
The sensitivity of ongoing searches for gravitational wave (GW) sources in the ultra-low frequency regime ($10^{-9}$~Hz to $10^{-7}$~Hz) using Pulsar Timing Arrays (PTAs) will continue to increase in the future as more well-timed pulsars are added to the arrays. It is expected that next-generation radio telescopes, namely, the Five-hundred-meter Aperture Spherical radio Telescope (FAST) and the Square Kilometer Array (SKA), will grow the 
number of well-timed pulsars to $O(10^3)$. The higher sensitivity will result in greater distance reach for GW sources, uncovering multiple resolvable GW sources in addition to an unresolved population. Data analysis techniques are, therefore, required that can search for and resolve multiple signals present simultaneously in PTA data. The multisource resolution problem in PTA data analysis poses a unique set of challenges such as non-uniformly sampled data, a large number of so-called pulsar phase parameters that arise from the inaccurately measured distances to the pulsars, and poor separation of signals in the Fourier domain due to a small number of cycles in the observed waveforms. We present a method that can address these challenges and demonstrate its performance on simulated data from
 PTAs with $10^2$ to $10^3$ pulsars. 
 The method estimates and subtracts sources from the data iteratively using multiple stages of refinement, followed by a step that mitigates spurious  identified sources by comparing the outputs from two different algorithms.
The performance of the method compares favorably with the global fit approaches that have been proposed so far. In all the cases tested in this work, the fraction of sources found by the method that correspond to true sources in the simulated data exceeds $78\%$ and $93\%$
for a large-scale (with $10^3$ pulsars and $200$ sources) and a mid-scale (with $10^2$ pulsars and $100$ sources) PTA, respectively. The network signal to noise ratio of the recovered true sources
reaches down to $16.43$ for the large-scale and $9.07$ for the mid-scale PTA.
\end{abstract}

\maketitle

\section{Introduction}\label{Sec:intro}
Gravitational wave (GW) astronomy is now well established in the high frequency ($10$ to $10^3$~Hz) band with more than 50 events detected so far by the Advanced LIGO and Advanced Virgo detectors~\cite{2016PhRvL.116f1102A, 2019PhRvX...9c1040A, 2020arXiv201014527A}. Work is underway to open up the $0.1$ to $10^2$~mHz band using space-based detectors: LISA~\cite{2017arXiv170200786A}, scheduled for launch around 2034, and its possible companions TianQin~\cite{2016CQGra..33c5010L}, Taiji~\cite{10.1093/nsr/nwx116}, DECIGO~\cite{2009JPhCS.154a2040S} and TianGo~\cite{2020PhRvD.102d3001K}. In the sub-$\mu$Hz band, pulsar timing arrays (PTAs) are putting increasingly tighter constraints on the stochastic GW background (SGWB) from an unresolved 
population of supermassive black hole binaries (SMBHBs) \citep{eptastochastic2015, 2015Sci...349.1522S, 2016ApJ...821...13A, 2018ApJ...859...47A}, continuous waves from resolvable SMBHBs \citep{2014MNRAS.444.3709Z, 2016MNRAS.455.1665B, 2019ApJ...880..116A}, and bursts \citep{2015MNRAS.446.1657W, 2020ApJ...889...38A}.

Based on its 12.5 yr data set with 47 millisecond pulsars (MSPs)~\cite{2021ApJS..252....4A, 2021ApJS..252....5A}, the North American Nanohertz Observatory for Gravitational Waves (NANOGrav) has claimed a discovery of a common-spectrum noise process with the Bayes factors exceeding  $10^4$~\cite{2020ApJ...905L..34A}. This signal
 could arise from an isotropic SGWB with a median characteristic GW strain amplitude of $1.92\times 10^{-15}$ for an $f^{-2/3}$ spectrum at $f=1/\rm{yr}$. However, the smoking gun evidence, i.e., the Hellings-Downs angular correlation \cite{1983ApJ...265L..39H}, is still lacking. 
 
The presence of a common-spectrum noise process is also strongly favored in the second data release from the Parkes Pulsar Timing Array (PPTA) collaboration~\cite{Goncharov_2021}. (The SGWB strain amplitude reported by the PPTA is within $1\sigma$ of that 
 obtained by NANOGrav.) The European Pulsar Timing Array (EPTA) also reports similar results~\cite{chenCommonredsignalAnalysis24yr2021} and the invesigation from International Pulsar Timing Array (IPTA)~\cite{antoniadisInternationalPulsarTiming2022} shows that these four results are in broad agreement with each other. However, as with the NANOGrav result, this data does not provide significant support for or against the  Hellings-Downs correlation. In addition, the PPTA  has investigated the possibility of model misspecification that can occur in PTA data interpretation. 
It is generally believed that additional data is required to decide if the observed common noise arises from GWs.

The sensitivity of PTA-based GW searches will get a significant boost over the next decade as  the Five-hundred-meter Aperture Spherical radio Telescope (FAST)~\citep{2011IJMPD..20..989N, 2019RAA....19...20H} and Square Kilometer Array (SKA)~\citep{2009A&A...493.1161S,2015aska.confE..37J} begin to grow the number of well-timed millisecond pulsars.
Pulsar surveys with FAST have already started and 201 new pulsars, including 40 MSPs, have been discovered in its recent Galactic Plane Pulsar Snapshot survey~\cite{2021RAA....21..107H}. The SKA will be constructed in two phases, i.e., SKA1 and SKA2. The former is currently under construction and it will contain more than $130,000$ dipole antennas in Western Australia and nearly $200$ parabolic dishes in South Africa. The latter will expand the number of antennas and dishes of SKA1 by a factor of ten, which represents the full science capability of the SKA observatory \footnote{SKA website:  \url{www.skatelescope.org}}. In this work, we consider a PTA based on anticipated SKA2 performance as this poses the strongest data analysis challenge.

In~\cite{2009A&A...493.1161S}, an SKA pulsar survey is simulated to estimate the number of normal ($\approx 14,000$) and millisecond pulsars ($\approx 6,000$) that would be detected. Observational strategies are then proposed by which timing of these pulsars for GW searches could be performed, resulting in an estimate of 6 to 20 hours for obtaining a single timing point for 250 millisecond pulsars at an individual SNR of 100. In~\cite{Hobbs_2019}, it is estimated that FAST will be able to time $50$ MSPs to a precision 
of $100$~ns in an observation period of $24$~h. Extrapolating from
these estimates, and using the anticipated upper bound of $100$~ns on the precision~\cite{2015aska.confE..37J,Hobbs_2019} in timing residuals, timing $\approx 10^3$ pulsars at this level of precision appears feasible. While the exact number of pulsars for future PTAs cannot be predicted, a PTA with $10^3$ pulsars timed to $100$~ns precision is a realistic target for the 
development of data analysis methods. This ensures that the science results in the SKA era will not be limited by inadequacies of data analysis methods but rather by instrumentation and observational limits.

Along with a more uniform sky coverage and standardized data spans, large-scale PTAs will improve the sensitivity to GWs from resolvable sources by two orders of magnitude~\citep{2017PhRvL.118o1104W,Erratum2017}. In tandem with an increase in sensitivity, the frequency range of PTA-based searches can stretch well beyond the conventionally assumed limit of $\sim 100$~nHz, set by the observation cadence for single pulsars, if appropriate asynchronous observation strategies are adopted~\cite{2021ApJ...907L..43W}. Among other things, this can bring the highly dynamical regimes of merger and ringdown in a binary coalescence under the purview of PTA-based GW astronomy.

Currently, dozens of SMBHB candidates (e.g., see  \cite{wangObservationalSignaturesClose2020,xinMultimessengerPulsarTiming2021} for brief reviews) in active galactic nuclei have been reported through electromagnetic signatures, especially periodic variations (e.g., \cite{valtonenMassiveBinaryBlackhole2008,grahamPossibleCloseSupermassive2015,liSPECTROSCOPICINDICATIONCENTIPARSEC2016,liPossibleSim202019}). There are emerging new techniques such as reverberation mapping \cite{wangKinematicSignaturesReverberation2018} and spectroastrometry \cite{songshengVLTInterferometricMeasurements2019} that can probe the dynamics of the gas close to an SMBHB directly and estimate the orbital parameters of the binary. This greatly enhances the opportunity of PTAs to resolve nHz GW sources with the orbital information of individual binaries \cite{arzoumanianMultimessengerGravitationalwaveSearches2020}.

Given the large distance reach of a future large-scale PTA for isolated SMBHBs -- for example, an SMBHB with a redshifted chirp mass of $4\times 10^{9}~M_{\odot}$ ($4\times~10^{10} M_{\odot}$) will be visible out to $z\simeq 1$ (28) (both with a rest frame chirp mass of a few $\times 10^{9}~M_{\odot}$) for an SKA-era PTA~\cite{2017PhRvL.118o1104W,Erratum2017}-- one expects to have not just one, but multiple SMBHB signals in the data. Disentangling and estimating the parameters of multiple resolvable sources against a noise background that includes the SGWB from unresolved sources is, therefore, an important data analysis challenge for large-scale PTAs.

While the multisource resolution problem is a subject of numerous studies in the context of space-based detectors, where several competitive approaches applicable to realistic data have been found~\cite{Littenberg:2011zg,Blaut:2009si,PhysRevD.104.024023}, it has received relatively little attention so far in the context of PTAs. In some ways, multisource resolution in PTA data presents a harder  set of problems: the timing residuals for individual pulsars are, in general, non-uniformly spaced in time; the observations of timing residuals are not temporally synchronized across the pulsar array; the typical data duration of $\sim 10$~yr encompasses only a few cycles of an SMBHB signal. The lack of uniform and synchronized sampling precludes the direct application of LISA multisource resolution methods that typically use Fourier domain techniques. The small number of cycles implies greater difficulty in distinguishing signals with nearby frequencies, resulting in a larger degree of confusion in multisource resolution.

In~\cite{PhysRevD.87.064036}, which builds on the earlier work in~\cite{PhysRevD.85.044034}, a global fit approach was explored for the multisource resolution problem in which, assuming a certain number of resolvable sources, a fitness function derived from the log-likelihood function of the data (the so-called \fstat) is maximized 
over the combined parameter space of the sources using a Genetic Algorithm.
A Bayesian alternative to the global fit approach in~\cite{becsy2020joint} uses the method
of reversible jump Markov chain Monte Carlo (RJMCMC) to obtain the best number of sources.
Several simplifying assumptions are made in the above works  to reduce the complexity of the multisource resolution problem. First, the number of resolvable GW sources and the number of array pulsars is limited to $\leq 8$ and $\leq 50$ in~\cite{PhysRevD.87.064036}, respectively, with the corresponding numbers being $3$ and $20$ in~\cite{becsy2020joint}. Secondly, while the GW signal in the timing residual for any one pulsar contains two contributions~\cite{2016MNRAS.461.1317Z}, called the Earth and pulsar terms, the latter is ignored in both approaches as it results in a simpler, albeit sub-optimal, fitness function. Some of these limitations are removed in~\cite{songshengSearchContinuousGravitationalwave2021}, where it is shown that the method of diffusive nested sampling (DNest) allows the Bayesian global fit approach to extend to $O(10^2)$ pulsars with both signal terms included.

In this paper, we present a method that uses an
iterative source extraction (\isis) approach, an alternative to global fit,
for the PTA multisource resolution problem. The \isis approach has been applied to LISA data~\cite{cornish2003lisa,Blaut:2009si,PhysRevD.104.024023}, showing performance comparable to global fit, and has been used  widely in astronomy, most notably 
in the CLEAN method~\cite{1974A&AS...15..417H}.
Our method has several refinements built on top of \isis such as a step where two semi-independent analyses are applied to the same data and the respective sets of identified sources are cross-checked against each other. A similar step was introduced in the context of the LISA multisource problem in~\cite{PhysRevD.104.024023} and found to be highly effective in eliminating spurious identified sources.

We call our method ``source analysis in pulsar timing array residuals with iterative swarm heuristics-based identification" (\algoname), 
since it uses Particle Swarm Optimization (PSO)~\cite{PSO,mohanty2018swarm} 
for the global maximization of the single-source fitness function. Using the 
single-source detection and parameter estimation methods proposed in~\cite{2015ApJ...815..125W,2017JPhCS.840a2058W}, \algoname takes into account both the Earth and pulsar terms in the GW signal. These methods can be scaled up to an arbitrary number of pulsars, allowing us to test multisource 
resolution for the unprecedented case of a PTA with $10^3$ pulsars. (We use the simulated PTA in~\cite{2017PhRvL.118o1104W} obtained from a realistic Galactic population distribution of MSPs~\cite{2009A&A...493.1161S}.) We test our method on data containing multiple independent realizations of a simple GW source population model. Each realization contains a much larger number ($100$ to $200$) of sources than earlier studies along with a very wide range of signal strengths. This provides a more realistic simulation of confusion noise from an unresolved population than the model of a purely Gaussian stochastic process, and it also tests the effect of spectral power leakage from extremely bright sources on the resolvability of weaker ones. 

The rest of the paper is organized as follows. In Sec.~\ref{Sec:single}, we describe how the parameters and signal waveform of a single monochromatic source  are estimated in PTA data. 
Sec. \ref{sec:mltsrcgen} discusses general considerations involved in an ISE approach to the multisource resolution problem for PTAs.
Sec.~\ref{Sec:multi} describes \algoname by first presenting the different key ideas and their justification, followed by the complete pipeline. The simulation setup used in this paper and the results obtained are described in Sec.~\ref{Sec:results}. This is followed by our conclusions in Sec.~\ref{Sec:conc}. 
\section{Single-source search}\label{Sec:single}

\subsection{Data model}
\label{sec:datamodel}

For a PTA consisted of $N_{\rm p}$ pulsars, our data to be analyzed is the set of the timing residual sequences, 
which is denoted as $\mathcal{Y} = \{\overline{y}^I\}$, $I = 1, 2,\ldots, \Np$. 
For each pulsar in an array, $\overline{y}^I\in \mathbb{R}^N$ is the sequence of the timing residuals observed at epochs $\{t^I_i\}$, $i = 1,2,\ldots,N$, where $\mathbb{R}^N$ is the space of $N$ element row vectors. 
We will use its associated continuous time notation, $y^I(t)$, instead of $\overline{y}^I$ when convenient. The elements of 
$\overline{y}^I$ are the differences between the measured pulse times of arrival (TOAs) and the predictions from the best-fit model (not including the effect of GWs) determined by parameters such as the sky position, proper motion, frequency, frequency derivatives of the pulsar, the dispersion measures at each observation epoch, and the Keplerian (and post-Keplerian for relativistic case) orbit parameters if the pulsar is in a binary system \citep{2006MNRAS.372.1549E}. 
The fitting procedure will, in principle, absorb the power of the signal that 
is degenerate with a timing parameter. This will decrease the sensitivity at 
corresponding frequencies \citep{1984JApA....5..369B}, such as at  
$1/({1~\rm yr})$ and $2/({1~\rm yr})$ due to the fitting of the sky position and parallax for the pulsar.

In general \cite{becsy2020joint},
\begin{align}
    \overline{y}^I &= \overline{s}^I(\theta) + \overline{n}^I  + \overline{e}^I \;,
\end{align}
where $\overline{s}^I(\theta)$, if present, is the GW-induced signal characterized by the set of parameters $\theta$, and  
$\overline{n}^I$ is a realization of the noise associated with the TOAs. The covariance matrix of the noise process is denoted by ${\bf C}^I$, with $C^I_{ij} = E[n^I_i n^I_j]$, $i, j = 1, 2, \ldots N$, and $n^I_i$ denoting the $i$-th element of $\overline{n}^I$.  
The error arising from the aforementioned fitting procedure is denoted by $\overline{e}^I$. It is typically expressed as $\overline{e}^I = M^{I} \delta p^{I}$, where  $ M^{I}$ is the design matrix used in the fitting procedure, and $\delta p^{I}$ denotes the differences between the best-fit and the true parameters \cite{2006MNRAS.372.1549E}.

The noise process $\overline{n}^I$ in timing residual data
is assumed to consist of three main components~\cite{2015JPhCS.610a2019W}: white noise rooted in radiometer and pulsar pulse profile  jitter; red noise arising from interstellar medium and spin irregularity of the pulsar~\cite{2010ApJ...725.1607S}; an additional red noise rooted in the stochastic gravitational-wave background.
In this paper we neglect the effects of red noise and timing model errors and assume that the samples of $\overline{n}^I$ are drawn from a zero mean white Gaussian noise process for which $C^I_{ij}=(\sigma^I)^2 \delta_{ij}$. 

Switching to the continuous time notation, the signal $s^I(t;\theta)$ induced by a GW source 
can be written as \citep{1975GReGr...6..439E, 2010PhRvD..81j4008S} 
\begin{equation}\label{eq:signal}
s^I(t;\theta) = \int_{0}^{t} \text{d}t' z^{I}(t';\theta)   \,,
\end{equation}
where $z^I(t;\theta)\equiv (\nu^I(t;\theta)-\nu^I_0)/\nu^I_0$ is the GW induced Doppler shift, and $\nu^I(t;\theta)$ and $\nu^I_0$ represent the spin frequencies of the pulsar observed at the Solar System Barycenter and at the pulsar, respectively. 
For a plane GW crossing the Earth-pulsar line of sight,  
emanating from a 
source located in equatorial coordinates at right ascension $\alpha$ and declination $\delta$, $z(t;\theta)$ can be expressed as 
\begin{equation}\label{eq:redshift}
z^I(t;\theta)  =  \sum_{A=+,\times} F^{I}_{A}(\alpha,\delta) \Delta h_{A} (t;\theta_s)   \,,
\end{equation}
where $\theta = \{\alpha, \delta\}\cup \theta_s$ and  $F^{I}_{+,\times}(\alpha,\delta)$ are the antenna pattern functions for the $+$ and $\times$ polarizations of the GW  \citep{2011MNRAS.414.3251L,2014ApJ...795...96W}. 
The two pulse response \citep{1975GReGr...6..439E},
\begin{equation}\label{eq:hpc}
\Delta h_{+,\times} (t;\theta_s)  =  h_{+,\times}(t;\theta_s)  -
h_{+,\times}(t - \kappa^I;\theta_s)  \,,
\end{equation}
contains the so-called Earth and pulsar terms that arise from the action of the GW on pulses at the times, $t$ and $t-\kappa^I$, of their  reception and emission, respectively.

For a non-evolving circular binary emitting a monochromatic signal, the parameter set $\theta_s$ contains
the overall amplitude ($\zeta$), GW frequency ($f_{\rm gw}$), 
inclination angle between the normal of the binary orbit plane and the line of sight ($\iota$), 
GW polarization angle ($\psi$), and initial orbital phase ($\varphi_0$) \citep{2014ApJ...795...96W}. 
The time delay $\kappa^I$ can be treated as an unknown constant phase offset, 
called the pulsar phase parameter $\phi_I$, for such a source \citep{2012ApJ...756..175E,2013CQGra..30v4004E}.

In the rest of the paper, it will be
convenient to switch from $\zeta$ to the network signal to noise ratio (SNR)
of a GW source as defined below:
\begin{align}
    {\rm SNR} & =  \left[\sum_{I=1}^{\Np}(\rho^I)^2\right]^{1/2}\;,\\
    \rho^I & = \| \overline{s}^I(\theta)\|^I\;,
\end{align}
where $\rho^I$ denotes the per-pulsar SNR and, for arbitrary sequences $\overline{x},\,\overline{y}\in \mathbb{R}^N$, $\|\overline{x}\|^I=[\langle \overline{x},\overline{x}\rangle^I]^{1/2}$, with
\begin{align}
    \langle \overline{x},\overline{y}\rangle^I & =
\frac{1}{(\sigma^I)^2}\sum_{i = 0}^{N-1} x_i y_i\;,
\end{align}
being the appropriate inner product that follows from the log-likelihood function under the assumed white Gaussian noise model~\cite{2015ApJ...815..125W}.


\subsection{Estimation of source parameters}
To search for resolvable GW sources, we 
use the likelihood based detection and parameter estimation methods
described in \citep{2015ApJ...815..125W,2017JPhCS.840a2058W,RAAPTR} 
that take both the Earth and pulsar terms into account. 
Following the commonly used
terminology in the GW literature, these methods treat
the pulsar phases as {\em extrinsic} parameters and the 
remaining ones as {\em intrinsic}. By definition, extrinsic parameters are those over which
the maximization or marginalization of the log-likelihood function can be performed using computationally efficient analytical or semi-analytical 
methods. On the other hand, intrinsic parameters are defined as those over which the maximization of 
the log-likelihood function requires the use of computationally expensive numerical methods. The choice of treating pulsar phase parameters as extrinsic allows the single-source methods above to scale to
an arbitrary $N_{\rm p}$.

The method in~\cite{2017JPhCS.840a2058W}, called AvPhase,
marginalizes over the pulsar 
phases, $\{\phi_I\}$, $I = 1,2,\ldots,N_{\rm p}$, semi-analytically.  The 
remaining parameters are estimated by maximizing the (marginalized)
likelihood using Particle Swarm Optimization~\citep{PSO,mohanty2018swarm,2016MNRAS.461.1317Z}.  
The maximum value thus obtained serves as the detection statistic for deciding between the
null ($H_0$) and alternative ($H_1$) hypotheses 
that a signal is absent or present, respectively, in the given data. The method in~\cite{2015ApJ...815..125W}, called MaxPhase,
is similar except the marginalization over $\{\phi_I\}$ is 
replaced by maximization.

Fixing the search range for PSO over the source frequency $f_{\rm gw}$ merits a brief discussion. It has been shown in~\cite{2021ApJ...907L..43W} that the conventional assumption that PTAs can only detect sources with $f_{\rm gw}< f_{\rm SP}$, the Nyquist rate corresponding to the average single pulsar observational cadence, is invalid in general. If the observations of the array pulsars are not synchronized, as is the case for real PTA data, it is possible to detect higher frequency signals that exist in the data due to aliasing. This is because the timing residual data for any one pulsar consists of discrete time samples of an underlying analog signal but these samples are acquired without using any anti-aliasing filter on the latter. A subtle point here is that not only is a higher frequency GW signal aliased in PTA observations but also noise. The nature of the latter may be elucidated by analyzing high cadence timing residual data \cite{2018MNRAS.478..218P, Dolch_2016, 2014MNRAS.445.1245Y}. We will ignore these subtleties in this paper and simulate PTA data, which will have temporally synchronized data samples across the array pulsars, under the assumption that there are no sources with $f_{\rm gw} > f_{\rm SP}$. This will also be the upper boundary of the PSO search range for $f_{\rm gw}$.

\subsection{Estimation of signal waveforms}
\label{sec:maxavphase}
In an \isis approach to multisource resolution, we need the  estimated 
signal waveform at each step. This 
means that we must estimate both the intrinsic as well as extrinsic parameters of a signal. Since the latter are marginalized in AvPhase and not estimated at all, AvPhase alone is not sufficient to carry out source subtraction. On the other hand, MaxPhase provides point estimates of all the parameters but has worse estimation error than AvPhase as the signal SNR becomes lower \cite{2017JPhCS.840a2058W}.

In \algoname, we combine AvPhase and MaxPhase in a mutually complementary way to circumvent the above issue. First, the intrinsic parameters of a source are estimated using AvPhase. Then the extrinsic parameter estimation step of MaxPhase is applied to the same data with the intrinsic parameters set to the AvPhase estimates. From here on, we refer to the combination of the two methods  as {\em MaxAvPhase}. 

\section{Multisource resolution: general considerations}
\label{sec:mltsrcgen}

To aid clarity in the discussions to follow, we use the following terminology
borrowed from~\cite{PhysRevD.104.024023}.
\begin{itemize}
    \item{\em Data}: When used in the context of the entire PTA, this term indicates the set  $\mathcal{Y} = \{\overline{y}^I\}$, $I = 1, 2,\ldots, \Np$ of timing residual sequences for all the array pulsars. 
    \item {\em Source}: When we refer to a GW source, we equate it to the parameters $\theta$ of the source where needed. Thus, the signal induced by this source in the timing 
    residual data, $y^I(t)$, of the $I$th pulsar is $s^I(t;\theta)$ and, by the subtraction of a source from the data, we mean $y^I(t) - s^I(t;\theta)$ for $I = 1, 2, \ldots, \Np$.
    \item {\em Identified sources}: The initial set of sources found by a multisource resolution method based on \isis. This set typically contains known spurious effects that must be 
    mitigated with further processing.
    \item {\em Reported sources}: The set of sources that is reported as the end product of a complete multisource resolution method.
    A complete method typically includes post-processing steps that eliminate identified sources which appear to be spurious.
    \item {\em Confirmed sources}: The subset of reported sources that match true sources as judged by some prescribed test of association. The test used in this paper is described in Sec.~\ref{sec:nmtc}. 
\end{itemize}
It is worth emphasizing here that in any analysis of real data, where we do not have prior knowledge of true sources, one would only have the sets of identified and reported sources as the output of a multisource resolution method. 

\subsection{Search space partitioning}
\label{sec:partition}
The sequence in which sources are extracted from data in the \isis approach
depends in a complicated way on the brightness of sources as well as their local density in the space of intrinsic parameters.
For example, it is possible that a source with a smaller SNR is identified before a stronger source if the former is in a more sparsely populated part of parameter space than the latter.
In addition, the chances of the identified SNR sequence becoming non-monotonic typically increases as the method digs for weaker sources in the data.

The lack of control on the properties of identified sources complicates the choice of the termination criterion for the iterations. In particular, if the
parameter region of the search 
is wide enough to have a significantly inhomogeneous source density, a termination criterion can be satisfied prematurely. Consider, for 
example, the criterion that stops the search if the SNR of an identified source falls below a 
preset threshold. If the search space contains both dense and sparse regions, such a criterion may get
satisfied by weaker sources in the sparse region, which are easier to find, before stronger sources in the dense region are identified. Shifting to a criterion that is independent of SNR, such as using a fixed number of iterations,
does not cure this problem either because  now the denser region could saturate this criterion before all resolvable sources in the sparser region are identified. 

For the above reasons, it is generally safer to split the search space into smaller regions within which 
source densities are more homogeneous (based on prior expectations about the source population distribution). For concreteness,  the natural parameter to partition  
in the case of LISA or PTA multisource resolution is the source frequency $f_{\rm gw}$, resulting in
\isis  carried out 
independently, and in parallel, in different frequency bands. 

\subsection{Metric for source association}
\label{sec:nmtc}

The sources identified or reported in multisource resolution will, in general, not have a 
one-to-one match, either in their parameters or waveforms, with any of the true sources in the data. Hence, a metric is required to quantify the closeness between two given sources, whether identified, reported, or true.  

A natural starting point for the metric, given two sources $\theta$ and 
$\theta^\prime$, is the cross-correlation coefficient of their signal waveforms,
\begin{align}
    R^I(\theta,\theta^\prime) & = \frac{C^I(\theta,\theta^\prime)}{\left[C^I(\theta,\theta)C^I(\theta^\prime,\theta^\prime)\right]^{1/2}}\;,\\
    C^I(\theta,\theta^\prime) & = \langle \overline{s}^I(\theta),\overline{s}^I(\theta^\prime)\rangle^I.
\end{align}
For an array of pulsars, a simple choice for the overall metric would be the average absolute value,
\begin{align}
    R_{\rm av}(\theta,\theta^\prime) &= (1/\Np)\sum_{I = 1}^{\Np} |R^I(\theta,\theta^\prime)|\;.
\end{align} 
However, in our tests, we found that this quantity does not have a reliable behavior when dealing
with $10^3$ pulsars. A possible reason for this is that $R^I(\theta,\theta^\prime)$ is insensitive to the SNRs of the two sources. This allows $|R^I(\theta,\theta^\prime)|$ to become high due to chance coincidence in, say, the frequencies of the two sources even if their SNRs are far apart. For such pairs of sources, therefore, $R_{\rm av}(\theta,\theta^\prime)$ can become high even though they are not matched well. The chances of this spurious coincidence happening, hence the chances of mis-association of sources, increase when considering an identified or reported source with a low SNR that is embedded in a higher density of sources.

With some
empirical testing, we arrived at a different way of combining the set of cross-correlation coefficients, described below, that resulted in more stable performance.  First, we define the auxiliary quantity
\begin{align}
    r(\theta,\theta^\prime) &= \frac{1}{\Np}n\left(\{I\,|\, |R^I(\theta,\theta^\prime)|\geq 0.9 \}\right)\;,
    \label{eq:nmtc}
\end{align}
where $n(A)$ is the number of elements in a set $A$. In words, $r(\theta,\theta^\prime)$ is the 
fraction of array pulsars for which the (absolute) correlation coefficient of the two signal waveforms exceeds some preset value. (As shown in Eq.~\ref{eq:nmtc}, we set this threshold at the ad hoc value of $0.9$ in this paper.) For a set of sources $\Theta = \{\theta_1,\theta_2,\ldots,\theta_M\}$ and a given source $\theta$, the
best match $\theta_{\rm max}\in \Theta$ to $\theta$ is
defined as,
\begin{align}
    \theta_{\rm max} & = \argmax_{\theta_i \in \Theta} \, r(\theta,\theta_i)\;.
\end{align}
Finding the best match source does not necessarily mean that the match is 
good. This is decided by
comparing $R_{\rm av}(\theta,\theta_{\rm max})$ to a user-specified threshold. 


\section{Multisource resolution: implementation details}\label{Sec:multi}

The core subroutine in \algoname, which we call {\em \visis}, is a straightforward iterative application of MaxAvPhase: If  $\theta_1$
to $\theta_{k-1}$ are the sources identified in the first $k-1$ iterations, the source in the next iteration, $\theta_k$, is obtained by 
running MaxAvPhase on the data residual $\mathcal{Y}_k = \{\overline{y}^I-\sum_{i=1}^{k-1} \overline{s}_i^I(\theta_i)\}$, 
$I = 1, 2, \ldots, \Np$. 
Following the discussion in Sec.~\ref{sec:partition}, the frequency search range is split into bands and \visis is applied 
to each one independently. The iterations are terminated once a preset number is completed. In the present version of \algoname, this number is kept the same for each frequency band.

In the remainder of this section, we first describe some elaborations of \visis along with the rationale behind them.
The complete \algoname pipeline, along with all its user-defined parameters, is presented at the end of the section. 

For convenience, a list of the acronyms referring to various components of \algoname is given below.
\begin{enumerate}
    \item ISE: iterative source extraction.
    \item SAPTARISHI: source analysis in pulsar timing array residuals with iterative swarm heuristics-based identification.
    \item MaxPhase: a single source detection algorithm by maximizing the log-likelihood function via maximizing over the pulsar phase, developed in \cite{2015ApJ...815..125W}. 
    \item AvPhase: a single source detection algorithm by maximizing the log-likelihood function via marginalizing over the pulsar phase, developed in \cite{2017JPhCS.840a2058W}. 
    \item MaxAvPhase: combination of AvPhase to estimate intrinsic source parameters,  followed by MaxPhase to estimate the extrinsic ones.
    \item \visis: ISE implemented using MaxAvPhase. 
    \item xBSE: crossband source elimination. 
    \item iBSE: inband source elimination. 
\end{enumerate}

\subsection{Crossband and inband source elimination}
\label{sec:xmblt}

As discussed earlier (c.f., Sec.~\ref{sec:partition}), the possibility of premature termination
suggests that single source searches 
should be carried out in frequency bands. However, the use of frequency bands leads to a new problem: it is typically observed that the number of spurious sources rises 
near the band edges due to the leakage of spectral power from bright sources in adjacent bands. In the case of LISA, a common strategy to mitigate this {\em edge effect} is to apply tapered and overlapping windows to the data in the Fourier domain~\cite{Crowder_2007,PhysRevD.104.024023}. Sources identified near the edges of a window are discarded, eliminating spurious sources, while genuine near-edge sources are recovered in adjacent overlapping windows. However,  Fourier domain techniques are not well-suited for PTA data as discussed earlier and this strategy, at least in a simple and direct form, cannot be ported over. 
While one can still
partition the search range of $f_{\rm gw}$ in MaxAvPhase into bands, a different approach is required
for the mitigation of edge effects.

The strategy used in \algoname for mitigating edge effects  is inspired by the method proposed in~\cite{mohantyMedianBasedLine2002} for removing instrumental line features (e.g., $60$~Hz and harmonics related to the North American electrical grid) in data from ground-based GW detectors. The method works in multiple stages. At first, using $l_1(t), l_2(t), \ldots, l_k(t)$ to denote the signals corresponding to the instrumental lines, $l_j(t)$ is estimated for each $j$ using a nonparametric method (based on the running median). In the next step, for each $j$, the estimated $l_i(t)$, for all $i\neq j$, are subtracted out from the data and $l_j(t)$ is
estimated again. Repeating these steps for all the lines
progressively reduces the estimation error in each $l_j(t)$ that arises from the mutual interference of the lines.

\algoname uses the above philosophy
in two forms, {\em crossband source elimination} (\xMBLT)
and {\em inband source elimination} (\iMBLT), as described below. The aim of \xMBLT is to reduce the edge effect while that of \iMBLT is the reduction of error in the estimated SNR of a source caused by the cumulative effect of weaker sources. This latter is observed to be substantial when there is a chance proximity of frequencies among true sources.

\xMBLT: Let $\mathcal{S}_m = \{ \theta_1^{(m)}, \theta_2^{(m)},\ldots, \theta_k^{(m)}\}$ be the set of identified 
sources  in frequency band $[f_m, f_m+ B_m]$ in one complete run of \visis. For each band $[f_m, f_m+ B_m]$, 
we subtract
all other identified sources, $\mathcal{S}_j$, $j\neq m$, from the data and run \visis on this band again to get a new set of identified sources. This process can be repeated multiple times, with $\mathcal{S}_m$
overwritten by the new set of identified sources at each stage, but we find that 
the convergence of estimates is rapid enough that using only a single stage works well. 


\iMBLT: We start with (i) a set  $\mathcal{S}_m=\{ \theta_1^{(m)}, \theta_2^{(m)},\ldots, \theta_k^{(m)}\}$ of identified sources
in band $[f_m, f_m+ B_m]$
sorted in descending order of SNR, and (ii) 
the residual, denoted by $\mathcal{R}_m$, after subtracting out from the 
data the  
sources, 
$\{ \theta_1^{(j)}, \theta_2^{(j)},\ldots, \theta_k^{(j)}\}$, $\forall j\neq m$, 
identified in all the other bands. In the next step,  $p\leq k-1$ sources other than the loudest, i.e., $\{ \theta_2^{(m)}, \theta_3^{(m)},\ldots, \theta_p^{(m)}\}$, are subtracted out
from $\mathcal{R}_m$ giving the residual $\mathcal{R}^\prime_m$. (Here, $p$ is 
a user-defined parameter of the \iMBLT algorithm.)  
A  new estimate of the loudest source, $\theta_1^{(m)}$, is obtained from $\mathcal{R}^\prime_m$ using MaxAvPhase.
The refined loudest source is subtracted out from $\mathcal{R}_m$ giving the residual $\mathcal{R}^{\prime\prime}_m$.  The
entire process repeats with $\mathcal{R}_m$ overwritten by $\mathcal{R}^{\prime\prime}_m$ and running \visis on (the new) $\mathcal{R}_m$. 
In the present version of \algoname, the number of \iMBLT iterations above is kept the same as the terminal number of iterations used in \visis.


\subsection{Automated frequency band selection}
\label{sec:bands}

All of the steps above, namely \visis, \xMBLT, and \iMBLT, start with a predefined set of frequency bands. However, the choice of these bands also affects the performance of these steps to some extent. For example,
if a true source happens to fall at or very close to a band edge, it will show up due to spectral power leakage as an identified source 
in both of the adjacent bands sharing this edge. In \xMBLT, when such a doubly identified source is removed, 
along with the other sources in a given band, it may also disappear in the adjacent band and, hence, not be found in either. To eliminate such spurious effects, band edges should ideally
be chosen to lie in sufficiently large gaps between the frequencies of resolvable sources.
The complication here is that the determination of bands should be based solely on estimated source frequencies since, in real data, there is no access to the true sources.

\begin{figure}
    \centering
    \includegraphics[scale=0.45]{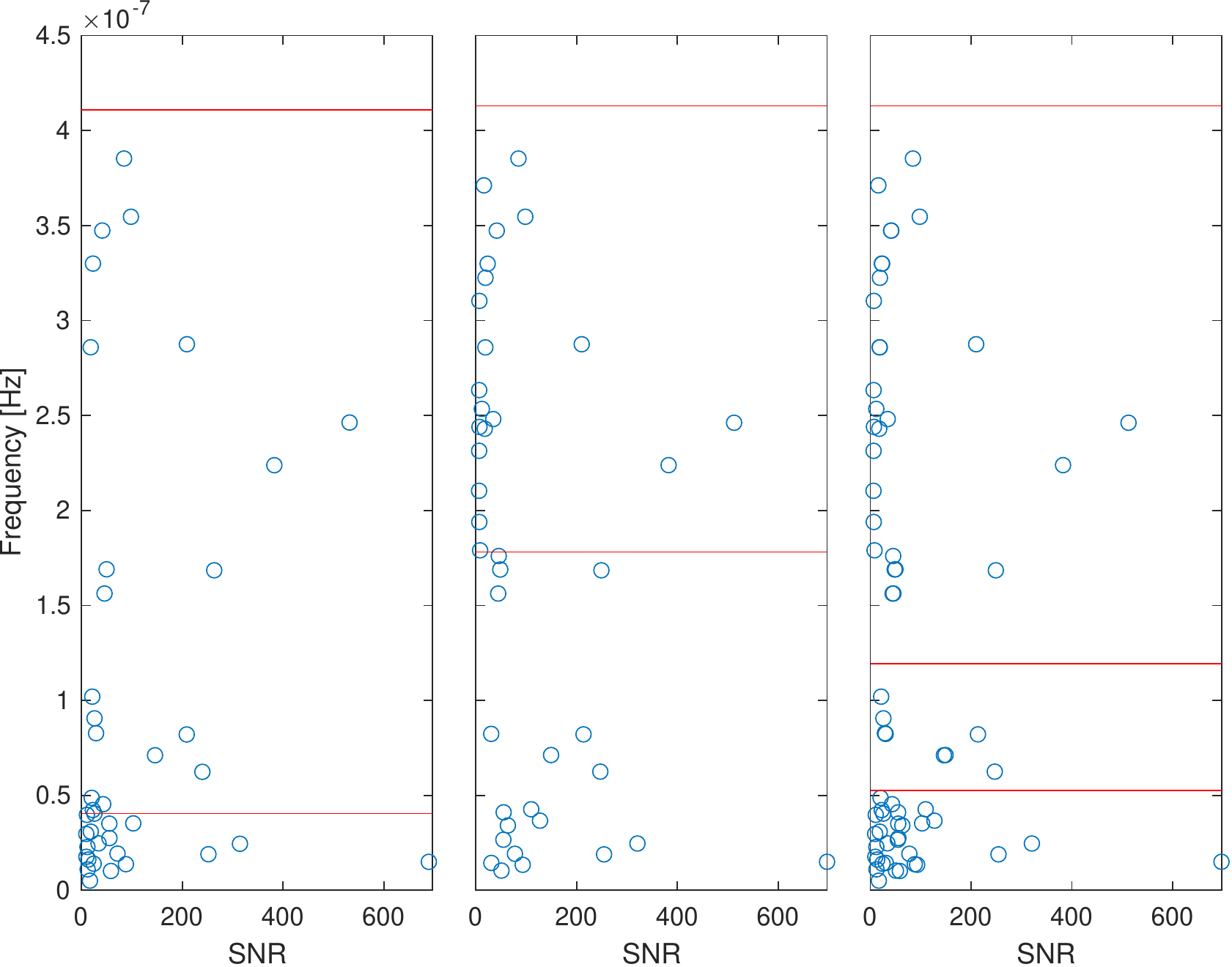}
    \caption{Illustration of the major steps in automated band selection. (The data realization used
    here is the first among the set described in Sec.~\ref{sec:sim}.) The left 
    and middle panels show sources identified by \xMBLT using two different sets of frequency band edges (red lines). The right panel shows the union of the two sets of sources after the merging of sufficiently close sources and the band edges (red lines)
    obtained from the second and third largest gaps in the sorted set of source frequencies.
    The horizontal
    and vertical axes show the estimated
    SNR and frequencies of the sources, respectively.}
    \label{fig:bandselect}
\end{figure}

Our algorithm for determining band edges, whose major steps are illustrated in Fig.~\ref{fig:bandselect}, starts with two different sets of arbitrarily chosen frequency bands, 
$\mathcal{B}_r = \{[0, f^{(r)}],[f^{(r)},f_{\rm SP}]\}$, $r=1,2$, $f^{(1)}\neq f^{(2)}$. For each $\mathcal{B}_r$, \xMBLT is used to find the set, $\Theta_r$,
of identified sources. For each  $\theta\in \Theta_1$, the best match source
$\theta^\prime \in  \Theta_2$ is found using the metric presented in Sec.~\ref{sec:nmtc}: if $R_{\rm av}(\theta,\theta^\prime) \geq \eta_{\rm band}$, where $\eta_{\rm band}$ is a user-defined parameter of \algoname, $\theta^\prime$ is replaced by $\theta$ in both sets. Next, the union
of the modified sets, $\Theta =\Theta_1\cup \Theta_2$, is obtained. Finally, the list of 
source frequencies in $\Theta$ is sorted in ascending order and the relative gaps between consecutive frequencies, $(f_{j+1}-f_j)/f_j$, is computed. The second and third largest relative gaps are found and the center of these two gaps are set as band edges. Thus, we 
get three frequency bands with this approach, which are found to be sufficient for analyzing
our simulated datasets. (We find that, for the source population distribution used in our simulations, discarding the largest gap is usually necessary in order to eliminate a 
very narrow starting band.) 

The basic reasoning behind the automated band selection
algorithm is that while using any one
$\mathcal{B}_r$ could lose on-edge sources in \xMBLT, taking the union of sources
found with multiple choices of $\mathcal{B}_r$ should recover these losses. Duplication of sources should be avoided in the union, hence the intermediate step above of merging sources
that are sufficiently close.


\subsection{Cross-validation}
\label{sec:cv}

One of the key components of \algoname is a step that we call {\em cross-validation}. The basic idea in cross-validation is that of comparing the identified sources obtained from two different algorithms in order to mitigate spurious ones. The details of the implementation are as follows.

Having obtained frequency bands as explained above, both \xMBLT and \iMBLT are
applied to the data using the same set of bands. (To save computational cost, \iMBLT is initialized using the sources identified with \xMBLT.) This yields two semi-independent sets of identified sources
$\Theta_{\rm \xMBLT}$ and $\Theta_{\rm \iMBLT}$ that, one expects, will have true sources in 
common but not spurious ones. 
A user-defined threshold $\eta_{\rm snr}$ is applied to both sets  such that identified sources with estimated ${\rm SNR}< \eta_{\rm snr}$ are discarded. (The modified sets continue to be denoted as $\Theta_{\rm \xMBLT}$ and $\Theta_{\rm \iMBLT}$.)

Next, for each source $\theta\in \Theta_{\rm \iMBLT}$,  the best match source $\theta^\prime \in  \Theta_{\rm \xMBLT}$ is found (c.f., Sec.~\ref{sec:nmtc}). If,
for a preset threshold  $\eta_{\rm cv}$, 
 $R_{\rm av}(\theta,\theta^\prime)< \eta_{\rm cv}$,
 $\theta$ is removed from $\Theta_{\rm \iMBLT}$. The final list of  the survivors in $\Theta_{\rm \iMBLT}$ constitute the set of reported sources
 from \algoname.

\begin{figure}
    \centering
    \includegraphics[scale=0.45]{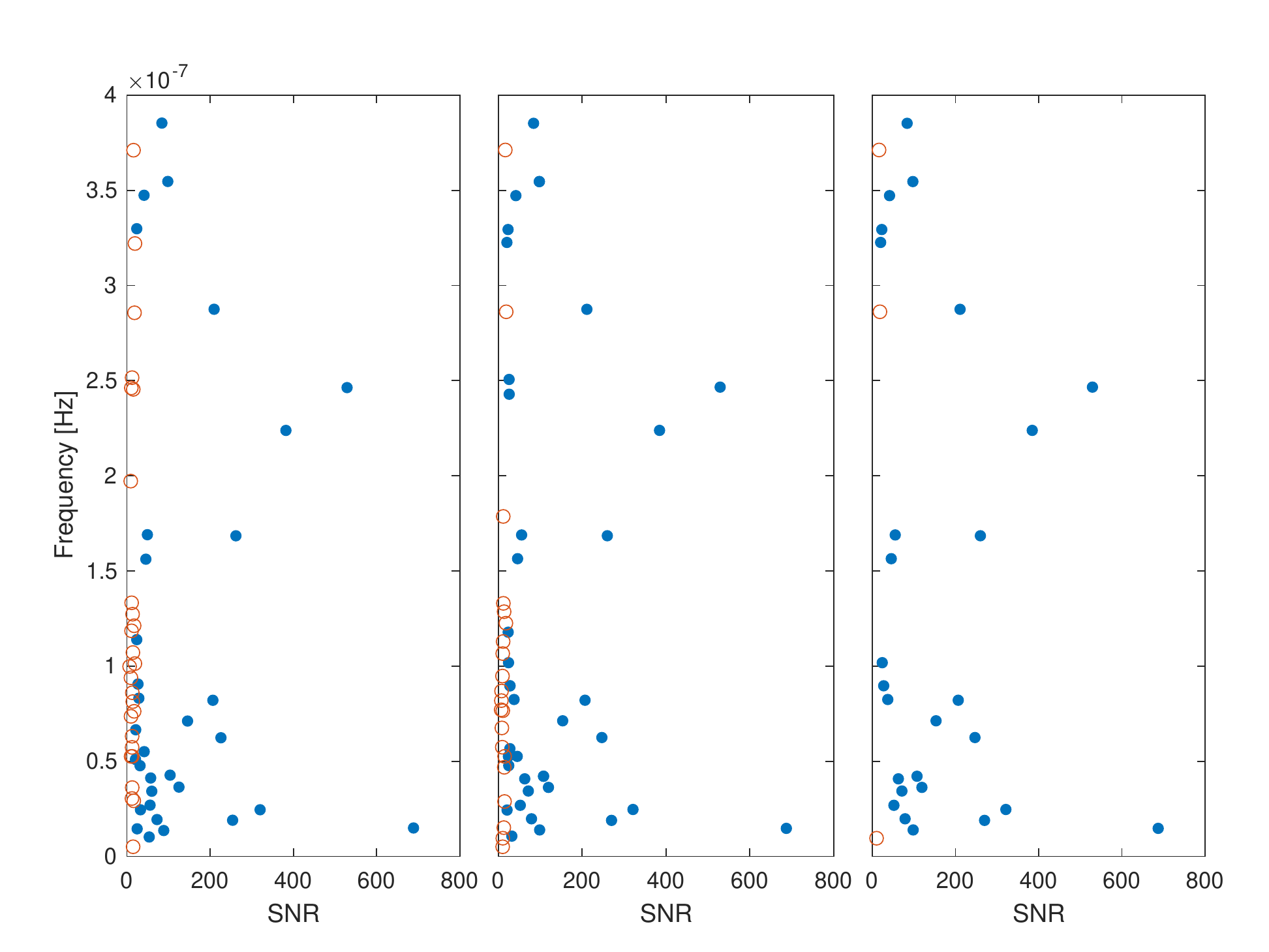}
    \caption{Illustration of cross-validation.  The left and middle panels show sources
    identified by \xMBLT and \iMBLT, respectively, while the right panel shows 
    the sources that survive after the cross-validation step. Open circles in each of the panels show sources with estimated ${\rm SNR} < 20$ while filled circles show the rest.  
    The horizontal and vertical axes show the estimated
    SNR and frequencies of the identified sources, respectively.}
    \label{fig:cvexample}
\end{figure}

Figure~\ref{fig:cvexample} illustrates the effect of cross-validation. 
Identified sources from \xMBLT and \iMBLT become progressively weaker in SNR as the iterations 
proceed. Eventually, the identified sources get affected by confusion noise arising from 
unresolvable sources and are more likely to be spurious in nature. As can be seen from the figure,
the estimated parameters of stronger sources show a good match between the two methods but the
weaker sources do not always coincide. (Here, for visual clarity, we have used an ad hoc boundary of ${\rm SNR}=20$ to separate identified sources into weak and strong.) Cross-validation leverages this observation to filter out spurious sources. This can be seen in the way that it predominantly targets the weaker sources for removal and spares most of the stronger ones.
 

\subsection{The \algoname pipeline}
\label{sec:pipeline}
We summarize here the complete \algoname pipeline here, starting with the list of its user-defined parameters.
\begin{itemize}
    \item $N_{\rm ise}$: The number of iterations of \visis in each frequency band.
    \item $p$: The number of sub-loudest sources subtracted in each step of \iMBLT.
    \item $\eta_{\rm band}$: The threshold used in automated frequency band selection for merging sufficiently close sources.
    \item $\eta_{\rm snr}$: The threshold on estimated ${\rm SNR}$ 
    used to discard identified sources prior to cross-validation.
    \item $\eta_{\rm cv}$: The threshold used in cross-validation to reject identified sources that do not have well-matched counterparts across \xMBLT and
    \iMBLT.
\end{itemize}
In addition to the above parameters are the ones for PSO.  In our experience with PSO in GW data analysis problems, we typically need to tune only two parameters in order to get good performance: the number of iterations ($\niter$) in one run of PSO, and the number of independent runs ($\nruns$) of PSO on a given optimization task. The rest of the parameters involved in PSO are described in \cite{2015ApJ...815..125W, wangCoherentMethodDetection2014} and are kept at the same values. We set $\niter = 2000$ and $\nruns = 8$ throughout this paper as this setting gives excellent performance while keeping computational costs manageable.

The steps involved in the \algoname pipeline are summarized below in sequential order.
\begin{enumerate}
    \item Automated band selection.-- This step is described in Sec.~\ref{sec:bands}. It requires two separate runs of \xMBLT with two different sets of frequency bands. Note that within each search band, the sources are identified using \visis.
    \item \xMBLT.-- This step uses the bands selected 
    in the previous step and follows the description in Sec.~\ref{sec:xmblt}.
    \item \iMBLT.-- The initialization of this step is carried out with the sources found in the previous step and follows the description in Sec.~\ref{sec:xmblt}.
    \item SNR threshold.-- Identified sources from both \xMBLT and \iMBLT that have an estimated SNR below $\eta_{\rm snr}$ are discarded. 
    \item Cross validation.-- This step, described in Sec.~\ref{sec:cv}, is carried out on the identified sources that survive the SNR threshold above.
   
\end{enumerate}
The final set of sources (i.e., reported sources) is the principal output of the pipeline.
Note that, as per the terminology in Sec.~\ref{sec:mltsrcgen}, this implies that the estimated 
values for all the signal parameters are obtained for each reported source.
For simulated PTA data, one
can proceed further to test for associations between the reported and true sources. 
This test simply follows the procedure described in Sec.~\ref{sec:nmtc}.

\section{Results}\label{Sec:results}

We begin this section with a description of our setup for simulating a large-scale PTA and its data. This is followed by the metrics we use to characterize the performance of \algoname and the presentation of our results.

\subsection{Simulation setup and \algoname settings}
\label{sec:sim}
We use two types of simulated PTAs, a mid-scale and a large-scale one, along with different numbers of GW sources to generate our simulated
data realizations. The MSPs in both PTAs are selected from the synthetic catalog provided in \cite{2009A&A...493.1161S} based on simulated SKA pulsar surveys: the large-scale PTA has $10^3$
pulsars within $3$~kpc from Earth; the mid-scale one has the closest $10^2$ pulsars in the large-scale PTA.

\begin{figure}
    \centering
    \includegraphics[scale=0.3]{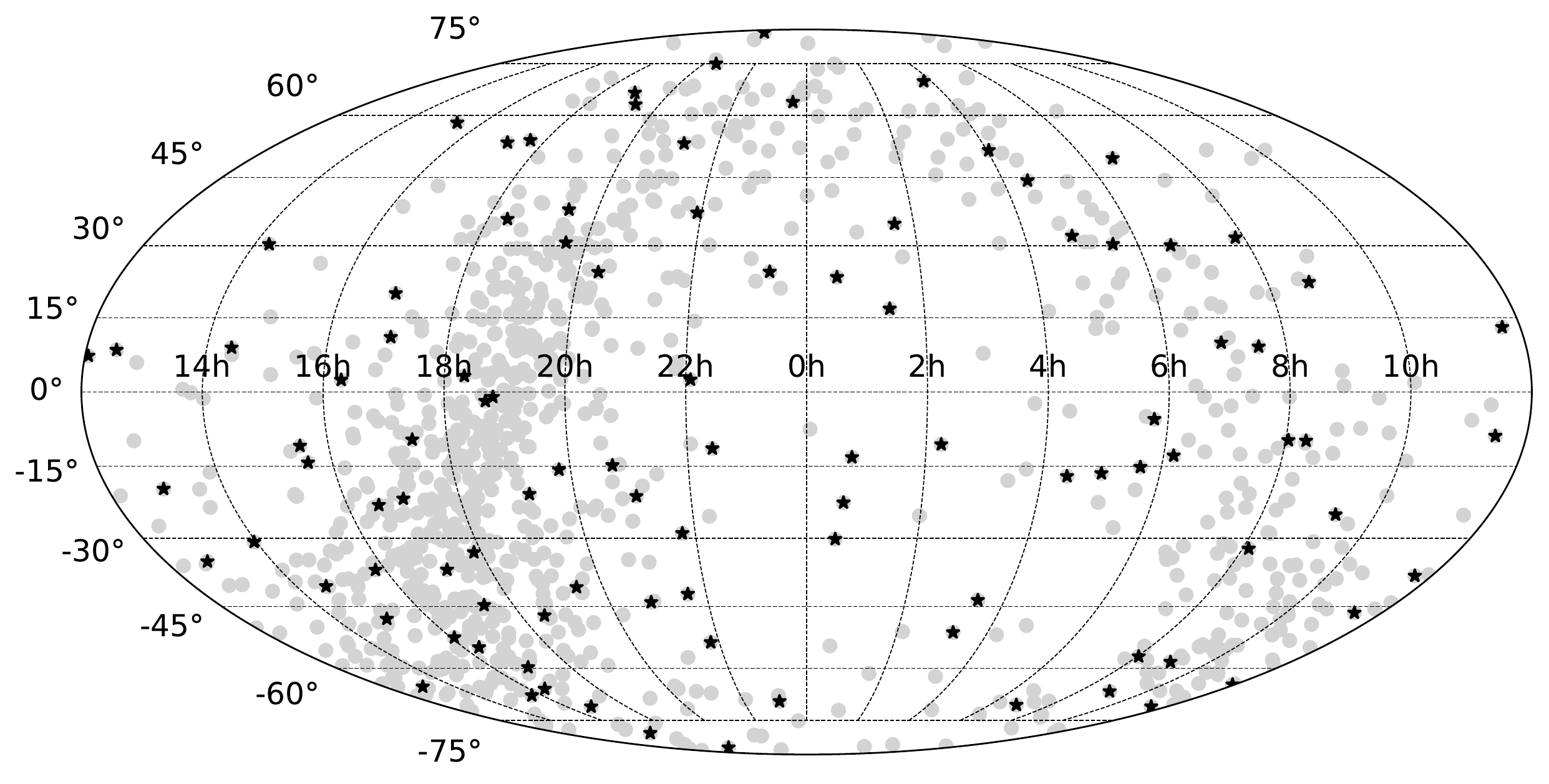}
    \caption{The sky distribution, in RA and Dec, of the pulsars in the
    simulated PTAs. The gray dots and black asterisks mark the locations of the pulsars in 
    the large and mid-scale PTAs respectively.}
    \label{fig:mspdist}
\end{figure}

Figure~\ref{fig:mspdist} shows the sky distribution of the pulsars in the above two PTAs. The 
main difference between the mid- and large-scale PTAs is the concentration of pulsars around the galactic plane. For the former, the pulsars are distributed more uniformly 
around the sky, which should lead to a different distribution of per-pulsar SNR values for
randomly placed GW sources.  

The timing residual for each pulsar is simulated following the data model in Sec.~\ref{sec:datamodel}. The noise standard deviation is set uniformly for all
pulsars to $100$~nsec, which is a conservative estimate of the timing error that is
expected in SKA-era observations. The 
observation times for all timing residuals are identical across a given PTA and 
have a cadence of 1~sample per $2$~weeks. The total duration of the data is $5$~yr. 

The number of GW sources for the large-scale PTA data is set to $200$ while it is $100$
for the mid-scale one. We generate $6$ independent data realizations, $\mathcal{Y}_1$ to $\mathcal{Y}_6$, as follows. Realizations
 $\mathcal{Y}_1$ to $\mathcal{Y}_5$ correspond to the large-scale PTA while $\mathcal{Y}_6$ corresponds to the mid-scale one. The latter is identical to the data used in \cite{songshengSearchContinuousGravitationalwave2021}.
  In all realizations,  the GW source parameters $\alpha \in [0,2\pi]$, $\cos\delta\in [-1,1]$, $\psi\in [0,\pi]$, $\varphi_0\in [0,2\pi]$, 
  $\log_{10}(f_{\rm gw})\in [0,\log_{10}(4.1\times 10^{-7})]$, and $\log_{10}(M)\in [6,10]$ ($M$ is the chirp mass of the SMBHB) are drawn independently from uniform distributions over their respective ranges. The distribution of the GW sources in distance corresponds to a constant spatial density up to a maximum of $10^3$~Mpc.
    In $\mathcal{Y}_1$ to $\mathcal{Y}_4$, $\iota$ is drawn from a uniform distribution on $[0,\pi]$, while in $\mathcal{Y}_5$ and $\mathcal{Y}_6$ the uniform distribution is over $\cos\iota \in [-1, 1]$. The different distributions of $\iota$  affect the SNR distributions of GW signals 
    by changing their polarization: $\iota = 0$ and $\pi/2$ correspond to circulary and lineary polarized signals, respectively, with intermediate polarizations for other values. This provides a simple diagnostic of the robustness of a multisource resolution method against variations in the SNR distribution.
    
    While the above distribution for the source population does not match the expected ones, such as \cite{sesanaSelfConsistentModel2010}, from realistic simulations, it is adequate for investigating the performance of multisource resolution methods. It has all the principal features, namely a  higher density of sources at lower frequencies and lower SNRs, that are required
    to stress test data analysis methods.

\begin{figure}
    \centering
    \includegraphics[scale=0.5]{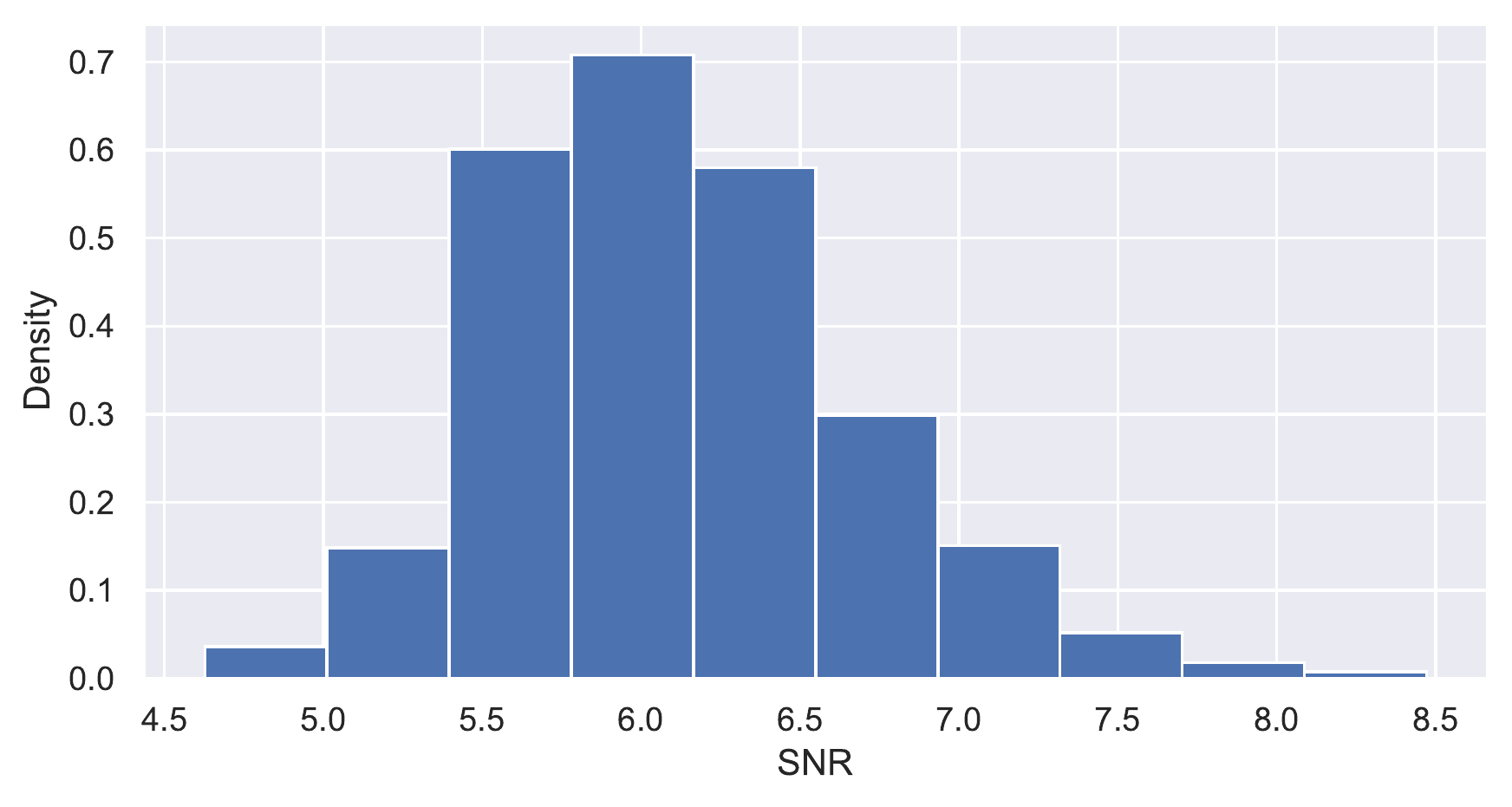}
    \caption{Distribution of SNR for identified sources obtained from noise-only data using \visis. To obtain this distribution, $50$ independent realizations of noise-only PTA data were analyzed with $20$ iterations of \visis per data realization. There was no division of the full frequency search range into smaller bands. This yields a total of $10^3$ identified sources in all.}
    \label{fig:snr_noise}
\end{figure}

The user-defined parameters of \algoname are listed in Sec.~\ref{sec:pipeline}. For the results presented here, we set  $N_{\rm ise}=20$, $p = 5$,  $\eta_{\rm band} = 0.8$, $\eta_{\rm snr} = 7.0 $, and $ \eta_{\rm cv} = 0.7$. 
The value chosen for $\eta_{\rm snr}$ is based on the SNR distribution of identified sources, as shown in Fig.~\ref{fig:snr_noise}, obtained using data realizations containing only noise.   
We set $\eta_{\rm snr} = 7.0$ as this is the closest round value to the $95$-th percentile   (${\rm SNR}=7.13$) of this distribution. One expects that most identified sources at or below this SNR would be spurious. 

\subsection{Performance metrics}
\label{sec:performance}
We use the following quantities to characterize the source resolution performance of \algoname on the simulated data described above. (i) {\em Detection rate.}--  Let the set of reported sources and the set of true sources be $\Theta_{\rm rep}$ and $\Theta_{\rm true}$, respectively. For each reported source $\theta \in \Theta_{\rm rep}$ the best match source $\theta^\prime\in \Theta_{\rm true}$ with ${\rm SNR}> 5.0$ is found. 
If, $R_{\rm av}(\theta,\theta^\prime)\geq \eta_{\rm conf}$, where $\eta_{\rm conf}$ is a preset threshold, $\theta$ is admitted into the set, $\Theta_{\rm conf}$, of confirmed sources. The detection rate is $n(\Theta_{\rm conf})/n(\Theta_{\rm rep})$ expressed as a percentage. (ii) {\em Lowest confirmed SNR.}-- This is the lowest estimated SNR in $\Theta_{\rm conf}$. We present results for 
 $\eta_{\rm conf}\in \{0.65, 0.70\}$ to characterize the variation in performance due to this user-defined parameter.

To quantify estimation performance, 
we consider differences in the values of signal parameters 
for matched pairs of confirmed and true sources. (While we will loosely refer to these differences as errors, it should be noted that they are not the same as the Frequentist or Bayesian notions of estimation error that always refer to a single true source.) 
In this paper, we restrict attention
to parameters of astrophysical importance, namely, SNR, $f_{\rm gw}$, RA($\alpha$) and Dec($\delta$). 

For SNR and $f_{\rm gw}$, we
use the absolute relative difference defined as 
$\Delta \chi = |\chi_{\rm conf}-\chi_{\rm true}|/\chi_{\rm true}$,  
with $\chi$ denoting one these parameters and the subscript on $\chi$ denoting the set $\Theta_{\rm conf}$ or $\Theta_{\rm true}$ that a source
belongs to. 
For the error in sky location, 
we use the shortest geodesic distance \cite{kellsPlaneSphericalTrigonometry}, $d_g$,  on the unit sphere  between a matching pair of confirmed and true source.
(We normalize $d_g$ by $2\pi$, the circumference  of a great circle on the unit sphere.) The set of errors for all confirmed sources is visualized as an empirical
{\em survival rate curve} $s(\epsilon)$. Here, $\epsilon$ is the estimation error ($\Delta \chi$ or $d_g/(2\pi)$)
as described above and $s(\epsilon)$ is
the fraction of confirmed sources that have a higher error than $\epsilon$.  

\subsection{Source resolution performance}
\label{sec:large_results}

\begin{figure*}
    \centering
    \includegraphics[scale=0.8]{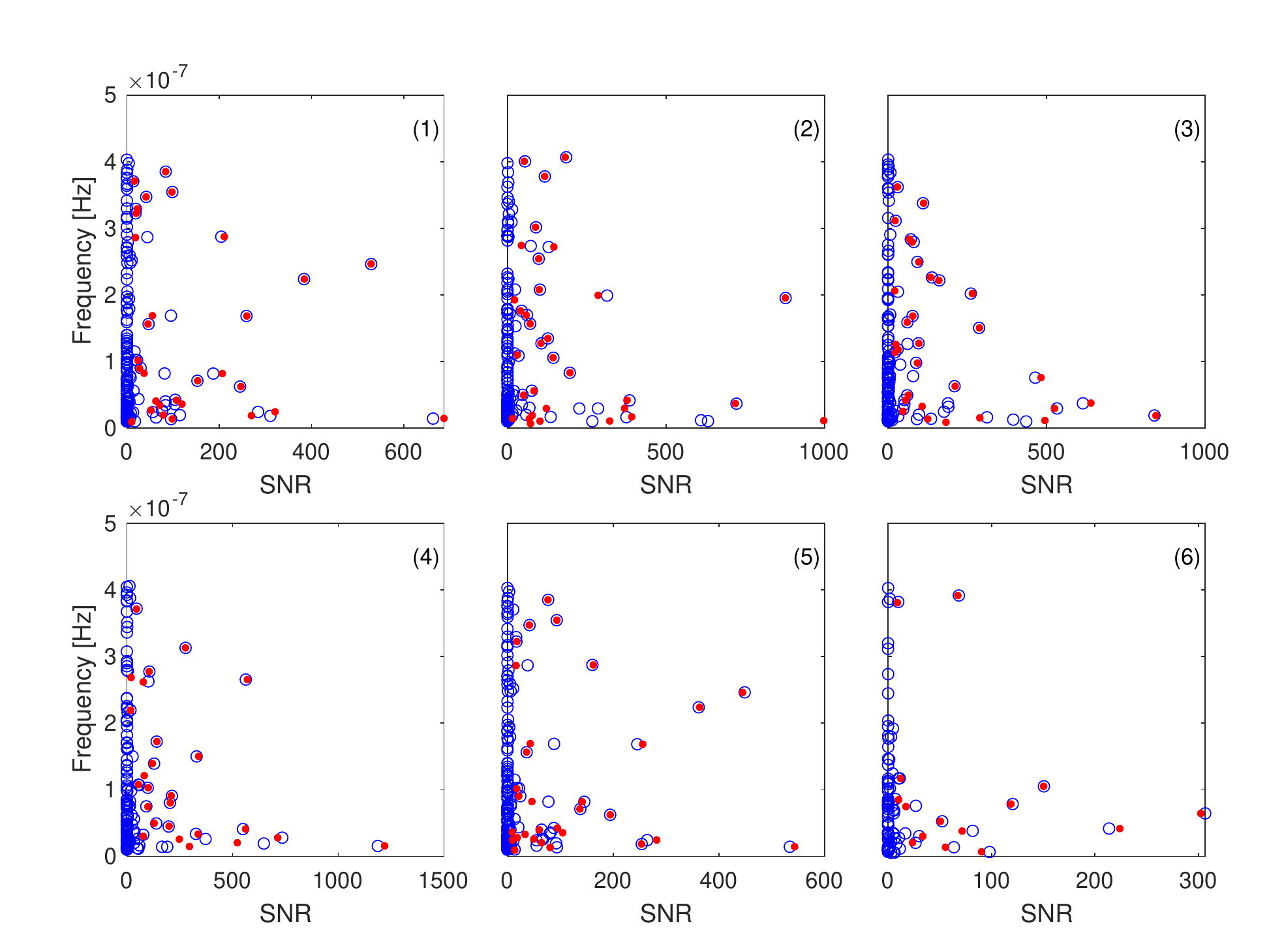}
    \caption{True (circles) and reported (dots) sources for the data realizations $\mathcal{Y}_1$ to $\mathcal{Y}_6$. The data realization corresponding to each panel is indicated by the number in parentheses at the top right corner. }
    \label{fig:srcres}
\end{figure*}

Figure~\ref{fig:srcres} shows the true sources used in all of the data realizations, $\mathcal{Y}_1$ to $\mathcal{Y}_6$, along with the reported sources from \algoname for each. 
We have chosen SNR and source frequency as the variables to display for each data realization. The former shows the large range of true source SNRs that was used in
each data realization while the latter shows that, in common with more realistic 
SMBHB population models, there is increasing crowding of sources towards lower frequencies. As a result, one expects that source resolution would worsen as one goes to lower frequencies. 
The source resolution performance of \algoname 
is summarized in Table~\ref{tab:srcres}. There are two parts to this table and they differ only
in the value of $\eta_{\rm conf}$ (c.f., Sec.~\ref{sec:performance}) used
in the confirmation of reported sources. As can be seen from the table, 
lowering $\eta_{\rm conf}$ naturally admits more of the reported sources into the set of confirmed ones, thereby increasing the detection rate and finding weaker sources. However, this also allows more spurious sources to get through. Note that confirming a weaker source with lower $\eta_{\rm conf}$ does not necessarily imply that its parameters are well-matched to its corresponding true source. This worsens parameter estimation performance, as is discussed later in Sec.~\ref{sec:mid_results}. In either case, we see that \algoname delivers good performance in source resolution: for the higher $\eta_{\rm conf}$ and more stringent cut on reported sources,  the detection rate stays above $78\%$, reaching its highest value of $93.3\%$ for $\mathcal{Y}_6$;  for 
the lower $\eta_{\rm conf}$, the detection rate is above $\simeq 85\%$ in all cases and reaches $100\%$ for $\mathcal{Y}_6$. 
\begin{table}[]
    \centering
    \begin{tabular}{|c|c|c|c|}
    \hline
    \multicolumn{4}{|c|}{(a) $\eta_{\rm conf}\geq 0.65$}\\
    \hline
    \shortstack{Data\\realization} & 
      \shortstack{No. reported \\ sources}  & \shortstack{Detection \\rate ($\%$)} &   \shortstack{Lowest  SNR\\ (Confirmed)}\\
      \hline
     $\mathcal{Y}_1$  & 30 & 96.7 & 16.50 \\
     $\mathcal{Y}_2$  & 33 & 84.8 & 31.04 \\
     $\mathcal{Y}_3$  & 31 & 93.5 & 21.05 \\
     $\mathcal{Y}_4$  & 26 & 88.5 & 18.28 \\
     $\mathcal{Y}_5$  & 32 & 90.6 & 8.86 \\
     $\mathcal{Y}_6$  & 15 & 100 & 9.07 \\
      \hline
    \multicolumn{4}{|c|}{(b) $\eta_{\rm conf}\geq 0.70$}\\
    \hline
    \shortstack{Data\\realization} & 
      \shortstack{No. reported \\ sources}  & \shortstack{Detection \\rate ($\%$)}&   \shortstack{Lowest  SNR\\ (Confirmed)}\\
      \hline
     $\mathcal{Y}_1$  & 30 & 86.7 & 18.64 \\
     $\mathcal{Y}_2$  & 33 & 78.8 & 31.04\\
     $\mathcal{Y}_3$  & 31 & 87.1 & 22.10\\
     $\mathcal{Y}_4$  & 26 & 84.6 & 18.28 \\
     $\mathcal{Y}_5$  & 32 & 78.1 & 16.43 \\
     $\mathcal{Y}_6$  & 15 & 93.3 & 9.07 \\
      \hline
    \end{tabular}
    \caption{Source resolution performance of \algoname. The two parts, (a) and 
    (b), correspond to different choices for the threshold $\eta_{\rm conf}$ used
    for confirming a given reported source.}
    \label{tab:srcres}
\end{table}

The results for $\mathcal{Y}_6$ allow a direct comparison between the \isis approach of \algoname and the global fit approach used in~\cite{songshengSearchContinuousGravitationalwave2021}. In the latter, the DNest method 
was able to resolve $8$ to $9$ sources with ${\rm SNR}\gtrsim 25$. From Table~\ref{tab:srcres} for $\eta_{\rm conf}=0.7$, we see that
\algoname resolves $14$ confirmed sources with the lowest 
${\rm SNR} = 9.07$. 

It is interesting to analyze the effect of cross-validation on source resolution 
performance. If this step is removed from the pipeline (c.f., Sec.~\ref{sec:pipeline}), we are
left with separate sets of reported sources from \xMBLT and \iMBLT  obtained by simply 
putting the threshold, $\eta_{\rm snr}\geq 7$, on the SNR of identified sources from each. 
Table~\ref{tab:srcres_solo} shows the results: Compared to part (b) of Table~\ref{tab:srcres}, there is a clear worsening of the detection rates, with steep drops for some of the data realizations such as $\mathcal{Y}_1$ and $\mathcal{Y}_4$. 

Sometimes high SNR sources are missed due to confusion arising from the chance clustering of multiple sources in frequency. This happens more at low frequencies where sources are more crowded. One way in which we have observed the manifestation of confusion is when a cluster of sources are close in both frequency and SNR such that they collectively mimic a spurious high SNR source. When this spurious source is estimated and removed from the data, the power of each constituent source in the cluster is also reduced, making it undetectable in the latter iterations. This mechanism is most clearly seen for data realization $\mathcal{Y}_2$ in Fig.~\ref{fig:srcres} where there is a spurious source at SNR $\approx 10^3$ that arises from the chance clustering of two sources with SNRs of $\approx 600$. The subtraction of the spurious source suppresses the detection of the latter two. The effect of confusion is smaller in $\mathcal{Y}_6$ 
due to the smaller number of sources. This explains why the detection rate is higher for this data realization. 
\begin{table}[]
    \centering
    \begin{tabular}{|c|c|c|c|}
    \hline
    \multicolumn{4}{|c|}{(a) \xMBLT}\\
    \hline
    \shortstack{Data\\realization} & 
      \shortstack{No. reported \\ sources}  & \shortstack{Detection \\rate ($\%$)} &   \shortstack{Lowest  SNR\\ (Confirmed)}\\
      \hline
     $\mathcal{Y}_1$ & 59 & 42.4 & 18.39  \\
     $\mathcal{Y}_2$ & 40 & 67.5 & 20.10 \\
     $\mathcal{Y}_3$ & 40 & 67.5 & 20.62 \\
     $\mathcal{Y}_4$ & 60 & 36.7 & 18.69  \\
     $\mathcal{Y}_5$ & 40 & 60 & 15.72  \\
     $\mathcal{Y}_6$ & 22 & 63.6 & 9.00  \\
      \hline
    \multicolumn{4}{|c|}{(b) \iMBLT}\\
    \hline
    \shortstack{Data\\realization} & 
      \shortstack{No. reported \\ sources}  & \shortstack{Detection \\rate ($\%$)}&   \shortstack{Lowest  SNR\\ (Confirmed)}\\
      \hline
     $\mathcal{Y}_1$  &  59  &  49.2  &  18.64  \\
     $\mathcal{Y}_2$  &  40  &  70  &  31.04  \\
     $\mathcal{Y}_3$  &  40  &  80  & 22.10   \\
     $\mathcal{Y}_4$  &  60  &  43.3  & 18.28   \\
     $\mathcal{Y}_5$  &  40  &  65  & 16.43   \\
     $\mathcal{Y}_6$  &  19  &  84.2  & 7.56   \\
      \hline
    \end{tabular}
    \caption{Source resolution performance without the cross-validation step. Reported sources are obtained separately from (a) \xMBLT and (b) \iMBLT using an SNR threshold $\eta_{\rm snr}\geq 7.0$ on identified sources. In both cases, the resulting reported sources are confirmed if $\eta_{\rm conf} \geq 0.7$.}
    \label{tab:srcres_solo}
\end{table}


\subsection{Parameter estimation performance}
\label{sec:mid_results}

\begin{figure}
    \centering
    \includegraphics[scale=0.6]{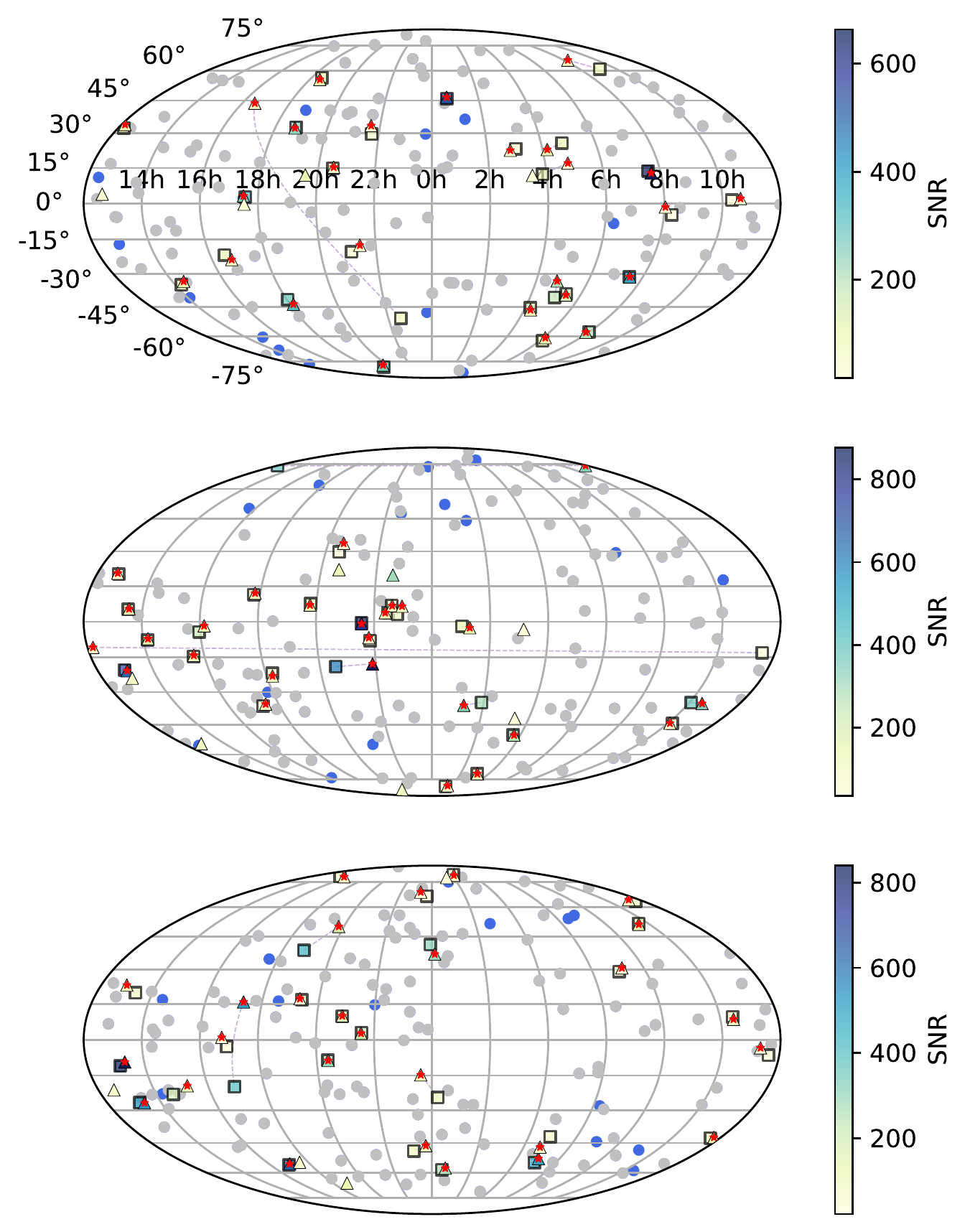}
    \caption{ The sky distribution, in RA and Dec, of true, reported, and confirmed sources for data realizations $\mathcal{Y}_1$ to 
    $\mathcal{Y}_3$ in order from top to bottom. In each plot, triangles show reported sources, stars represent confirmed sources, squares show true sources that matched a confirmed source, blue filled circles are true sources above the lowest confirmed SNR that were not detected, and filled gray circles show all other true sources. There is a dashed line joining each matching pair of confirmed and true sources but, the localization error being small in general, these lines are clearly visible only in a few instances. There is a long dashed line in the panel for $\mathcal{Y}_2$ but it actually connects a pair of sources that are quite close on the sphere. The triangle and square markers are 
    shaded according to the estimated and true SNRs, respectively. The colorbar on the right of each panel shows the correspondence between the colors and SNRs. 
    \label{fig:molwd_1_3}}
\end{figure}

\begin{figure}
    \centering
    \includegraphics[scale=0.6]{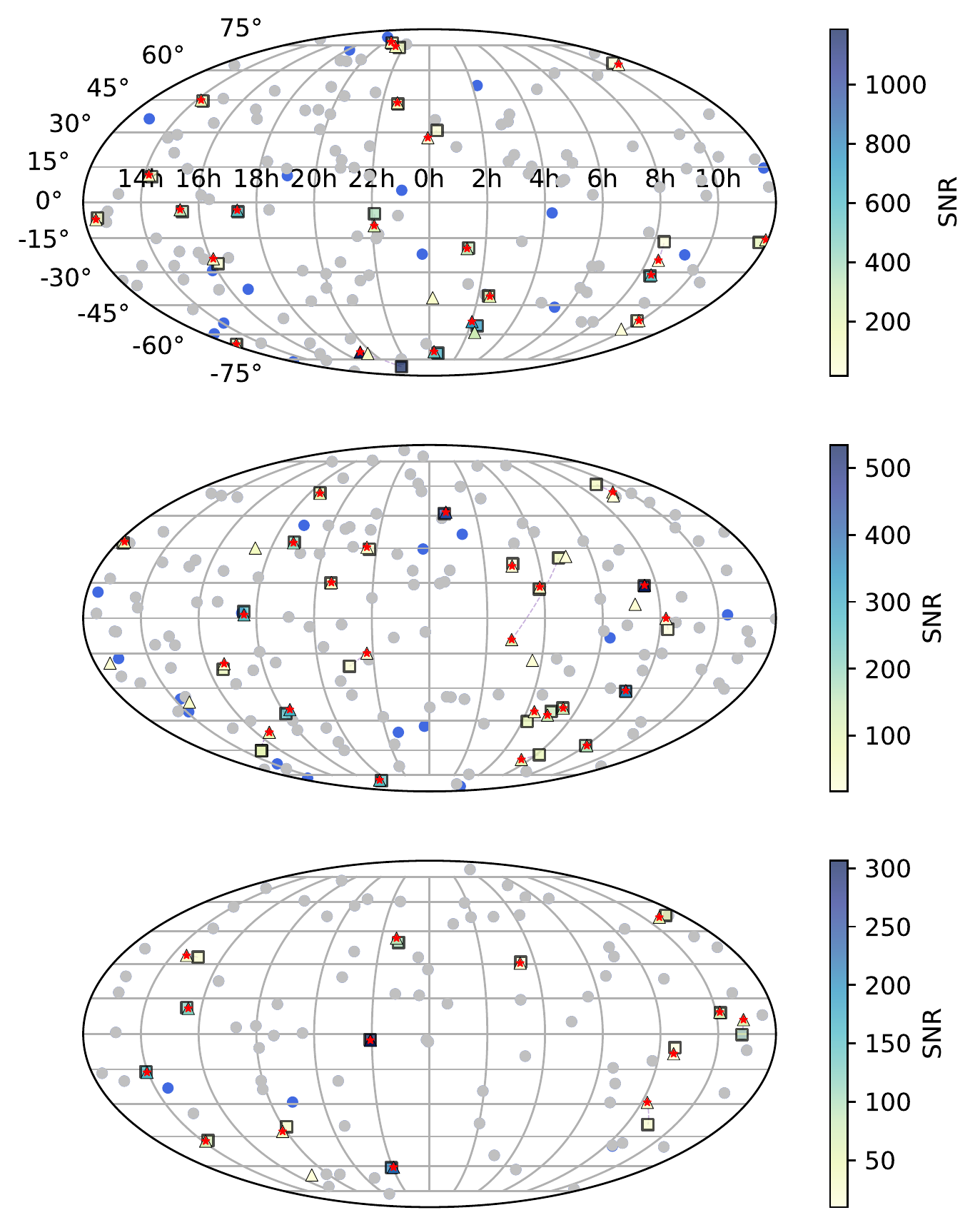}
    \caption{The sky distribution, in RA and Dec, of true, reported, and confirmed sources for data realizations $\mathcal{Y}_4$ to 
    $\mathcal{Y}_6$ in order from top to bottom. The description of markers, colors, etc., is the same as for Fig.~\ref{fig:molwd_1_3}.}
    \label{fig:molwd_4_6}
\end{figure}

Figures~\ref{fig:molwd_1_3} and~\ref{fig:molwd_4_6} show the sky distributions of true and reported sources for all the data realizations. 
For these plots, the reported sources correspond to part (b) of Table~\ref{tab:srcres} with the threshold for elevating a reported source to confirmed set at the more stringent
value of $\eta_{\rm conf}=0.7$. It is evident that the localization errors are small in general, even at lower SNRs. This is highly encouraging for the
prospects of multi-messenger 
astronomy with large-scale PTAs in the scenario where SMBHBs are first identified using gravitational 
waves and then followed up in the electromagnetic window.

\begin{figure*}
    \centering
    \includegraphics[scale=0.4]{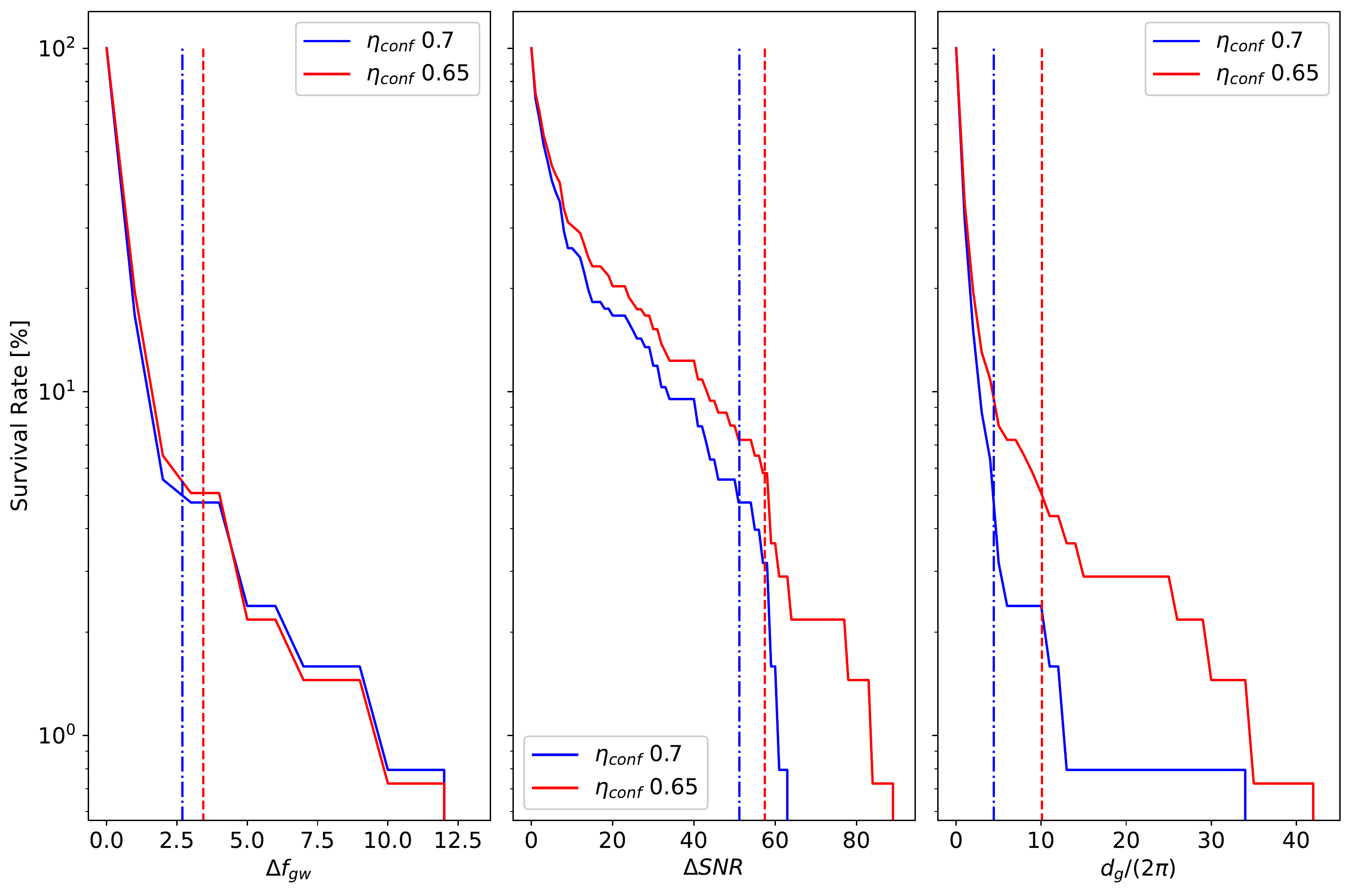}
    \caption{Survival rate curves for the errors (expressed as percentage) in $f_{\rm gw}$, SNR, 
    and sky localization. The latter is quantified using the normalized shortest geodesic distance $d_g/(2\pi)$. The curves in blue and red correspond to different choices for $\eta_{\rm conf}$, the threshold used to elevate a reported source to a confirmed one. Corresponding to each curve, a vertical dashed line of the same color shows the $95$th percentile
    of the error distribution. Their specific values are as follows. For $\eta_{\rm conf} = 0.7$,  $\Delta f_{\rm gw} = 2.7\%$, $\Delta {\rm SNR} = 51.19\%$, and $d_g /(2\pi) = 4.425\%$. For $\eta_{\rm conf} = 0.65$, $\Delta f_{\rm gw} = 3.44\%$, $\Delta {\rm SNR}=57.46\%$, and $d_g/(2\pi) = 10.1\%$.}
    \label{fig:surv_rate}
\end{figure*}

The errors in sky localization, SNR, and $f_{\rm gw}$ are further 
quantified in Fig.~\ref{fig:surv_rate}  using survival rate curves as described in Sec.~\ref{sec:performance}. Here, 
we only consider data
realizations $\mathcal{Y}_1$ to $\mathcal{Y}_5$ since $\mathcal{Y}_6$ is drawn from a very different model. Instead, we provide 
summary statistics for it in Table~\ref{tab:y_6_param_est}. In Fig.~\ref{fig:surv_rate}, the principal survival curves correspond to $\eta_{\rm conf} = 0.7$ but we also 
show the ones for  $\eta_{\rm conf}=0.65$ to simply illustrate the point discussed earlier lowering
$\eta_{\rm conf}$ increases detection rates but worsens parameter estimation.
\begin{table}[]
    \centering
    \begin{tabular}{|c|c|c|c|}
    \hline
     & $\Delta {\rm SNR}$  & $\Delta f_{\rm gw}$ & $d_g/(2\pi)$ \\
     \hline
     Mean   & 8.71\% & 0.77\%  & 0.87\% \\
     Minimum & 0.19\%  & 0 & 0.09\%  \\
     Maximum & 35.77\% & 5.09\% & 3.62\%  \\
     \hline
    \end{tabular}
    \caption{Summary statistics for the parameter estimation performance of \algoname on data realization $\mathcal{Y}_6$. Each column shows the sample mean, 
    minimum and maximum values of the errors, as defined in Sec.~\ref{sec:performance}, 
    for the parameter indicated in the heading of that column.}
    \label{tab:y_6_param_est}
\end{table}

Not surprisingly, and in line with Fisher information analysis,
$f_{\rm gw}$ is the most precisely determined parameter 
because it is the one that the likelihood function is most sensitive to. While a 
large fraction of sources are localized to within $10\%$ of a complete great circle 
emanating from a true source, there are a few large outliers. The number of these outliers
grows as the threshold for confirmation, $\eta_{\rm conf}$, is lowered because more spurious reported sources manage to enter the set of confirmed sources.

The parameter estimation performance of \algoname for $\mathcal{Y}_6$ (c.f., Table~\ref{tab:y_6_param_est}) compares
well with the global fit approach in~\cite{songshengSearchContinuousGravitationalwave2021}. Here, it should be emphasized that the notion of error is 
different in frequentist and Bayesian methods: while \algoname 
provides a point estimate of parameters, with the inference of errors requiring a separate analysis using multiple data realizations, the DNest method provides 
samples from the joint (or marginalized) posterior probability density function of the parameters for a single data realization. Nonetheless, a comparison of our results for $\mathcal{Y}_6$ in
Fig.~\ref{fig:molwd_4_6} with Fig.~4 of~\cite{songshengSearchContinuousGravitationalwave2021}
shows that the errors in sky localization are similar in size. 

\subsection{Computational considerations}
\label{sec:compcost}
The current code for \algoname is a mix of Matlab and C language codes. For computational efficiency, the core routines, which includes MaxAvPhase, are all in the latter language. The overall computational cost is dominated by the 
 \visis runs required in automated frequency band selection, \xMBLT and \iMBLT. Essentially all of the cost of \visis resides in the PSO-based optimization required in MaxAvPhase.

The runtime of \algoname is reduced considerably by exploiting the parallelism inherent to many of its components. The runs of \visis on separate frequency bands are carried out in parallel on different compute nodes of a cluster. The multiple PSO runs required in MaxAvPhase execute in parallel on the processing cores of a single node. As a result, the runtime for \visis on a given frequency band 
is simply the 
time required to execute MaxAvPhase with one PSO run -- about 1 hour for the 
large-scale PTA
on an Intel Xeon Platinum 8160 processor with a $2.1$~GHz clock rate -- multiplied by the number of sources to be identified.  Given that we have 
set the latter at $20$ in this paper, it takes about 20 hours for one \visis
run to finish on one frequency band. The total time taken to do one complete run of \algoname on one data realization then depends on how many \visis runs are used in total. 

At present, however, the number of \visis runs is not a realistic predictor of the total time for a complete run. This is because the 
current structure of the \algoname code requires multiple jobs to run sequentially on a shared computing resource and each job incurs an unpredictable wait time in the queue. As we did not separately track the wait times,  we can only report here that it typically takes about a week at present to analyze one data realization completely.

There is considerable scope for vastly speeding up \algoname that has not  been tapped yet. Each iteration of PSO uses multiple independent evaluations of the function to be optimized and 
parallelizing over these can lead to a substantial reduction (about a factor of $\simeq 40$) in runtime. It is also possible in the case of MaxAvPhase to implement
parallelization within a single function evaluation with hardware-level acceleration using Graphics Processing Units (GPUs) or multi-threaded numerical libraries. This could easily bring in another factor of $O(10)$.  We plan to implement these nested layers of parallelization in the future. (Substantially speeding up the code will also help mitigate the problem of wait times if the complete run could be executed as a single job within the maximum time allowed for a job.)

\section{Conclusions}\label{Sec:conc}
 
We have introduced \algoname, a method for resolving SMBHB sources
in data from PTAs. The method uses an \isis approach and performs well relative to the global fit methods that have been proposed for the PTA multisource resolution problem. It is scalable to PTAs with a large number of pulsars and its application to
simulated data demonstrates that the
source confusion problem for future large-scale PTAs is solvable.

\algoname was applied to two types of PTAs, a large-scale one 
with $10^3$ pulsars and a mid-scale one with $10^2$ pulsars. 
Detection rates above $\simeq 78\%$ were achieved in all cases. 
The lowest confirmed SNR for the large and mid-scale PTAs
reached $16.43$ and $9.07$, respectively. Note that,
assuming a naive $\sqrt{N_{\rm p}}$ scaling of SNR with the number of pulsars, the source with
the lowest confirmed SNR for the large-scale PTA 
would have an SNR of only $\simeq 5.2$ for the mid-scale one, making it too weak to resolve with the latter. This indicates the usefulness of a larger number of array pulsars for not only improving single source search sensitivity~\cite{2017PhRvL.118o1104W} but also multisource resolvability.

The levels of lowest confirmed SNR above bode well for the ability
of a future large-scale PTA to peer far back into the Universe.
For example, using the lowest confirmed SNR of $16.43$  as a reference and an observation period of 5 yr, a SMBHB with $f_{\rm gw}=2 \times 10^{-8}$~Hz and redshifted chirp mass of $10^{9}~M_{\odot}$ ($10^{10}~M_{\odot}$) will be located at a redshift of $z= 0.24$ ($z= 5.84$) or a luminosity distance (standard cosmology) of $1.23$~Gpc ($57.3$~Gpc). [The corresponding chirp mass in the source frame is $8\times 10^{8}~M_{\odot}$ ($1.5\times 10^{9}~M_{\odot}$).] Here, we have used the relation in~\cite{2017PhRvL.118o1104W} between redshift, or distance, and the SNR 
averaged over the geometrical factors that occur in the response of the large-scale PTA.

The successes of both the \isis and global fit approaches suggest
that the strengths of the two can be combined to create an 
approach that is more powerful. For example, a weakness of the \isis
approach is that the proximity of two sources in frequency 
leads to an overestimate, due to their overlap, of the louder SNR and an underestimate of the weaker one when the louder is subtracted out. On the other hand, as shown in  \cite{PhysRevD.85.044034},
a global fit approach can estimate some source parameters accurately even if the sources coincide in frequency. 

Similarly, a weakness of the global fit approach is the necessity of assuming the number of sources and then using some form of model selection to find the best number. Finding the global fit becomes harder as the dimensionality of the combined parameter space increases with the number of sources, requiring special measures to combat this effect. An \isis approach does not need the number of sources and can identify as many as required. As shown in this paper, post-processing steps such as cross-validation can then be used to weed out a large fraction of spurious sources. It is conceivable that the number of reported sources from the \isis approach provides a good initial estimate for a global fit approach.

The synergy of the \isis and global fit approaches described above, as well as additional ones, will be explored in future work. Code modifications outlined in the paper to make \algoname substantially faster are in progress. This will allow its detection and estimation performance to be quantified more thoroughly using a much larger set of data realizations. 

A limitation of the current code is the use of the white noise model. While appropriate for the development of data analysis methods, effects such as red noise and timing model errors (c.f., Sec.~\ref{sec:datamodel}) need to be included in order to analyze real data. The key components of SAPTARISHI, namely xBSE, iBSE, and cross-validation, use the estimated source waveforms and parameters produced by the single-source estimation method, MaxAvPhase, as inputs. Thus, these steps are independent of the details of the single source method, allowing it to be substituted easily with any other method in the future that incorporates a more sophisticated noise model.

{\em Red noise:} Under the likely scenario that individual timing residuals are noise dominated, the covariance matrix of the noise, including red noise, can be estimated from the data. This estimate can then be fed into the log-likelihood. Alternatively, a parametrized model for the covariance matrix can be used and the parameters estimated jointly with those of the signal. Since both MaxPhase and AvPhase are derived from the log-likelihood, the modifications required are, in principle, straightforward and the computational burden is likely to increase only modestly. 

{\em Timing model errors:} For incorporating timing model errors, the standard approach uses a linear model with unknown coefficients~\cite{1984JApA....5..369B}. In general, the log-likelihood for Gaussian noise is a convex function over linear parameters. We expect the log-likelihood maximized or marginalized over the pulsar phase parameters to retain a similar simple structure over these parameters. Therefore, it is likely that with a suitable nested maximization scheme, in which the timing model error parameters are maximized over first, the optimization by PSO of the log-likelihood over the larger set of intrinsic parameters remains feasible albeit with a higher computational cost. Further work is in progress on this topic.

\acknowledgments
Y. W. gratefully acknowledges support from the National Natural Science 
Foundation of China (NSFC) under Grants No. 11973024 and No. 11690021, Guangdong Major Project of Basic and Applied Basic Research (Grant No. 2019B030302001) and the National Key Research and Development Program of China (No. 2020YFC2201400). We acknowledge the Texas Advanced Computing Center (TACC) at the University of Texas at Austin (www.tacc.utexas.edu) for providing high performance computing resources. 
The authors acknowledge the helpful discussion on the global fit approach using diffusive nested sampling and SMBHB astrophysics with Yu-Yang Songsheng and Prof. Jian-Min Wang at Institute of High Energy Physics at Chinese Academy of Science. We thank the anonymous referees for helpful comments and suggestions.

\appendix
\section{AvPhase}\label{Sec:App}
Following \cite{2015ApJ...815..125W}, for a PTA of $\Np$ pulsars, the log-likelihood ratio is
\begin{eqnarray}
\ln\Lambda(\mathcal{Y};\lambda) & = & \sum_{I=1}^{\Np} \ln\Lambda_I({\overline{y}^I};\lambda)\,,
\label{eq:loglambda}
\end{eqnarray}
where for the $I$-th pulsar,  
\begin{equation}\label{eq:loglambda2}
    \begin{aligned}
    \ln \Lambda_{I}(\overline{y}^I ; \lambda) &=\left[\left\langle \overline{y}^I \mid X_{I}\right\rangle_{I} \cos 2 \phi_{I}+\left\langle \overline{y}^I \mid Y_{I}\right\rangle_{I} \sin 2 \phi_{I}\right.\\ 
    &+\left\langle \overline{y}^I \mid Z_{I}\right\rangle_{I} -\frac{1}{2}\left(\left\langle X_{I} \mid X_{I}\right\rangle_{I} \cos ^{2} 2 \phi_{I}\right.\\
    &+\left\langle Y_{I} \mid Y_{I}\right\rangle_{I} \sin ^{2} 2 \phi_{I}+2\left\langle X_{I} \mid Y_{I}\right\rangle_{I} \sin 2 \phi_{I} \cos 2 \phi_{I}\\
    &+2\left\langle X_{I} \mid Z_{I}\right\rangle_{I} \cos 2 \phi_{I}+2\left\langle Y_{I} \mid Z_{I}\right\rangle_{I} \sin 2 \phi_{I}\\
    &\left.\left.+\left\langle Z_{I} \mid Z_{I}\right\rangle_{I}\right)\right] .
    \end{aligned}
\end{equation}
Here the explicit expressions of $X_{I}$, $Y_{I}$ and $Z_{I}$ can be found in \cite{2015ApJ...815..125W}. 

AvPhase marginalizes the likelihood ratio over the pulsar phase parameters (c.f., Sec.~\ref{sec:datamodel}) and maximizes the resulting function over the remaining parameters. This creates a new detection statistic, which we call Marginalized-Maximized Likelihood Ratio Test, \cite{2017JPhCS.840a2058W} given by
\begin{equation}\label{eq:maxie}
\text{MMLRT}(\mathcal{Y})=\max_{\lambda_i}\ln\left(\marg_{\lambda_e} \Lambda(\mathcal{Y};\lambda)\right) \,,
\end{equation}
where {\it marg} denotes the marginalization operation.
Here, the maximization over intrinsic parameters $\lambda_i$ is handled by PSO while the marginalization over the extrinsic parameters $\lambda_e$ is handled by numerical integration. 

The marginalization over the set of pulsar phase parameters, $\{\phi_I\}$, $I = 1,2,\ldots,\Np$, in Eq.~\ref{eq:maxie} is similar in spirit to existing proposals in Bayesian approaches to SMBHB search but differs significantly in implementation.  In~\cite{2013CQGra..30v4004E}, marginalization over $\{\phi_I\}$ (and other parameters) is performed on the estimated posterior probability density function (PDF), obtained with an MCMC based method, defined on the full parameter space of a single source. Note that this is not the same as sampling a posterior  that is already marginalized over $\{\phi_I\}$ since the dimensionality of the parameter space explored by MCMC is vastly different in the two cases.  
The substantial computational cost of MCMC based estimation of the posterior PDF defined over the full parameter space makes this approach ill-suited for scaling to SKA-era PTAs.  
In~\cite{2014PhRvD..90j4028T}, a solution was proposed in the form of direct numerical integration of the posterior over $\{\phi_I\}$, carried out at each iteration of the MCMC method. While this reduces the computational cost, the time to convergence of the numerical integration method reportedly varies a lot with the SNR of the injected signal since it affects the shape of the posterior in the $\Np$-dimensional space of $\{\phi_I\}$. In AvPhase, the marginalization of the likelihood over pulsar phases is performed economically using $\Np$ decoupled one-dimensional integrations that, as described below, are carefully crafted to avoid numerical issues and incur a  (nearly) fixed computational cost.

\subsection{Marginalization over pulsar phases}

To compute MMLRT one can rewrite the log-likelihood ratio function, i.e. Eq.~\ref{eq:loglambda2}, in the following form:
\begin{equation}
    \begin{aligned}
\ln\Lambda_I(\overline{y}^I;\lambda)  =& b_1 +  b_2\cos2\phi_{I} 
+b_3\sin2\phi_{I} +\\
& b_4 \sin2\phi_{I}\cos2\phi_{I} +b_5\cos ^{2} 2\phi_{I}
+\\
& b_6\sin^{2} 2\phi_{I}  \,. 
\end{aligned}
\label{eq:loglambda3}
\end{equation}
Here, $b_1 = \langle \overline{y}^I \mid Z_{I} \rangle_{I} - \frac{1}{2} \langle Z_{I} \mid Z_{I} \rangle_{I}$, 
$b_2 = \langle \overline{y}^I \mid X_{I} \rangle_{I} - \langle X_{I} \mid Z_{I} \rangle_{I}$, 
$b_3 = \langle \overline{y}^I \mid Y_{I} \rangle_{I} - \langle Y_{I} \mid Z_{I} \rangle_{I}$, 
$b_4 = -\langle X_{I} \mid Y_{I} \rangle_{I}$, 
$b_5 = -\frac{1}{2} \langle X_{I} \mid X_{I} \rangle_{I}$, and
$b_6 = -\frac{1}{2} \langle Y_{I} \mid Y_{I} \rangle_{I}$.  
Note that we have suppressed the index $I$ of $b$'s for clarity. 
Thus, the likelihood ratio function $\Lambda_I=\exp(\ln \Lambda_I)$ is 
\begin{equation}
    \begin{aligned}
\Lambda_I(\overline{y}^I;\lambda)  &= \exp(b_1 +  b_2\cos2\phi_{I} 
+b_3\sin2\phi_{I} \\
&+ b_4 \sin2\phi_{I}\cos2\phi_{I} +b_5\cos ^{2} 2\phi_{I}
+b_6\sin^{2} 2\phi_{I})  \,. 
\end{aligned}
\label{eq:loglambda4}
\end{equation}
The marginalization of $\Lambda$ over the pulsar phases involves $\Np$ 
decoupled integrations as follows 
\begin{equation}
    \begin{aligned}
\Lambda_I(\overline{y}^I;\lambda_i)  &= \int_{0}^{\pi} \exp(b_1 +  b_2\cos2\phi_{I} 
+b_3\sin2\phi_{I} \\
&+ b_4 \sin2\phi_{I}\cos2\phi_{I}+b_5\cos ^{2} 2\phi_{I}\\
&+b_6\sin^{2} 2\phi_{I})~\text{d}\phi_{I}  \,. 
\end{aligned}
\label{eq:int1}
\end{equation}
Set $x=2\phi_{I}$, then 
\begin{equation}
    \begin{aligned}
\Lambda_I(\overline{y}^I;\lambda_i)  &= \frac{1}{2}\int_{0}^{2\pi} \exp(b_1 +  b_2\cos x 
+b_3\sin x \\
&+ b_4 \sin x\cos x +b_5\cos ^{2} x+b_6\sin^{2} x)~\text{d} x  \,. 
    \end{aligned}
\label{eq:int2}
\end{equation}
Eq.~\ref{eq:int2} is equivalent to the marginalization of the posterior PDF in the 
Bayesian framework with a flat prior over $x$.  For the special case $b_4 = b_5 = b_6 =0$,  the integration 
has a closed form solution in terms of the modified Bessel function of the first kind.
In general though, Eq.~\ref{eq:int2} does not have a closed form 
solution and numerical integration must be used. While this appears straightforward,  it turns out that the power of the exponential in the integrand can occasionally become very large (say $>10^{8}$), 
making the integrand exceed the dynamical range of floating point precision to become
practically infinite. However, since what we really want is the logarithm of 
Eq.~\ref{eq:int2}, we can mitigate this problem by factoring out the dominant part of the integrand as explained next.
\subsection{Integration}
One can regard $b$'s as the coefficients of the basis functions, i.e. ($1, \sin x, \cos x, \sin x\cos x, \cos^2 x, \sin^2 x$), 
which have ranges from -1 to 1. The largest absolute value of the terms will determine 
the factorization of the integration. And we note that the basis functions change their 
sign in each of the four quadrants of $[0, 2\pi)$. To find out which term is dominating, 
we rewrite the integration as follows 
\begin{equation}
    \begin{aligned}
\Lambda_I(\overline{y}^I;\lambda_i)  &= \frac{1}{2}\sum_{k=1}^{4} \int_{l_k}^{u_k} \exp(b_1 S_{k1}^2 
+  b_2 S_{k2}^2\cos x \\
&+b_3 S_{k3}^2 \sin x + b_4 S_{k4}^2 \sin x\cos x \\
&+b_5 S_{k5}^2\cos ^{2} x+b_6 S_{k6}^2\sin^{2} x)~\text{d} x  \,. 
\end{aligned}
\label{eq:int21}
\end{equation}
Here $S_{kl}$ is the element of the sign matrix 
\begin{equation}\label{sign}
\mathbf{S}=\left( \begin{array}{cccccc}
1 & 1 & 1 & 1 & 1 & 1 \\
1 & -1 & 1 & -1 & 1 & 1 \\
1 & -1 & -1 & 1 & 1 & 1 \\
1 & 1 & -1 & -1 & 1 & 1 \\
\end{array} \right)\,.
\end{equation}
$S_{kl}^2=1$. Thus, we can redefine the new coefficients in quadrant $k$ as $b_{sl}^k=S_{kl} b_l$, and the new basis functions in quadrant $k$ as $S_{kl} V_l$ which are always positive in each quadrant. (Note, the summation rule is not applied here.)

Therefore, Eq.~\ref{eq:int21} can be explicitly written as 
\begin{subequations}
\begin{align}
\Lambda_I(\overline{y}^I;\lambda_i)  &= \frac{1}{2}\int_{0}^{2\pi} \exp(f(x))~\text{d} x \,, \\
&=\frac{1}{2}\sum_{k=1}^{4}e^{6N_k}\int_{l_k}^{u_k} g_k(x)~\text{d} x  \,, \\
&=\frac{1}{2}\sum_{k=1}^{4}e^{6N_k}\exp(\ln(R_k))\,.
\label{eq:int3}
\end{align}
\end{subequations}
where $N_k=\max(b_s^k)$, $R_{k}=\int_{l_k}^{u_k} g_k(x)~\text{d} x$. 
In quadrant $k$, the integration can be written as 
\begin{subequations}
\begin{align}
\int_{l_k}^{u_k} \exp(f(x))~\text{d} x &=\int_{l_k}^{u_k} \exp(b_1 S_{k1}^2 
+  b_2 S_{k2}^2\cos x \nonumber\\
&+b_3 S_{k3}^2 \sin x + b_4 S_{k4}^2 \sin x\cos x \nonumber\\
&+b_5 S_{k5}^2\cos ^{2} x +b_6 S_{k6}^2\sin^{2} x)~\text{d} x  \,,  \\
&=e^{6N_k}\int_{l_k}^{u_k} \exp(b_{s1}^k +  b_{s2}^k |\cos x| \nonumber\\
&+b_{s3}^k |\sin x| + b_{s4}^k |\sin 2x\cos 2x| \nonumber\\
&+b_{s5}^k |\cos ^{2} x|+b_{s6}^k |\sin^{2} x| -6N_k) ~\text{d} x  \,.  
\label{eq:int22}
\end{align}
\end{subequations}
The exponent in Eq.~\ref{eq:int22} is well behaved (after subtracting $6N_k$, each 
term in the exponent should be less than one), therefore it can be integrated 
using the usual algorithms without special treatment. 

To sum the contributions from the four quadrants, we define $M_k =6N_{k}+\ln R_{k}$, then 
\begin{equation}
\Lambda_I(\overline{y}^I;\lambda_i)  =\frac{1}{2}\sum_{k=1}^{4}\exp(M_{k})  \,.
\label{eq:int4}
\end{equation}
Set $\bar{M}=\max(M_k)$, and $\tilde{M}_k=M_k-\bar{M}$, 
\begin{equation}
\Lambda_I(\overline{y}^I;\lambda_i)  = \frac{1}{2}\exp(\bar{M}) \sum_{k=1}^{4}\exp(\tilde{M}_{k})  \,,
\label{eq:int4}
\end{equation}
\begin{equation}
\ln\Lambda_I(\overline{y}^I;\lambda_i)  = \bar{M} + Q - \ln 2\,,
\label{eq:llr}
\end{equation}
where $Q=\sum_{k=1}^{4}\exp(\tilde{M}_{k})$.

\bibliography{bibliography.bib}

\end{document}